\documentclass[acmlarge,screen, nonacm]{acmart}

\renewcommand\footnotetextcopyrightpermission[1]{} 

\usepackage{tabularx}
\usepackage{multirow}
\usepackage{booktabs}
\usepackage{array}
\usepackage{balance}  
\usepackage{graphics} 
\usepackage{mathptmx}
\usepackage{multicol}
\usepackage{float}
\usepackage{soul}
\usepackage{enumitem}
\usepackage{todonotes}
\usepackage{mdframed}
\usepackage{xcolor}
\usepackage[normalem]{ulem}
\usepackage{graphicx}
\usepackage{subcaption}

\AtBeginDocument{%
  }

\newtoggle{revisions}
\togglefalse{revisions}
\iftoggle{revisions} {
\newcommand{\rev}[1]{{\color{blue}{#1}\normalfont}}
}{
  \newcommand {\rev}[1]{#1} 
 }

\begin{document}

\title{I Can’t Join, But I Will Send My Agent: Stand-in Enhanced Asynchronous Meetings (SEAM)}

\author{Zhongyi Bai}
\affiliation{%
 \institution{School of Computer Science, The University of Sydney}
 \streetaddress{Camperdown/Darlington}
 \city{Sydney}
 \state{New South Wales}
 \country{Australia}
 \postcode{2008}}
\email{zhongyi.bai@sydney.edu.au}

\author{Jens Emil Grønbæk}
\affiliation{%
 \institution{Department of Computer Science, Aarhus University}
 \city{Aarhus}
 \country{Denmark}}
\email{jensemil@cs.au.dk}

\author{Andrew Irlitti}
\affiliation{%
 \institution{School of Information Technology, Deakin University}
 \city{Geelong}
 \country{Australia}}
\email{andrew.irlitti@deakin.edu.au}

\author{Jarrod Knibbe}
\affiliation{%
 \institution{School of Electrical Engineering and Computer Science, The University of Queensland}
 \city{St Lucia}
 \state{QLD}
 \country{Australia}}
\email{j.knibbe@uq.edu.au}

\author{Eduardo Velloso}
\affiliation{%
 \institution{School of Computer Science, The University of Sydney}
 \streetaddress{Camperdown/Darlington}
 \city{Sydney}
 \state{New South Wales}
 \country{Australia}
 \postcode{2008}}
\email{eduardo.velloso@sydney.edu.au}

\renewcommand{\shortauthors}{Bai et al.}

\begin{abstract}
We propose and explore the user experience of SEAM---Stand-in Enhanced Asynchronous Meetings---virtual reality meetings in which embodied virtual agents represent absent users. During the meeting, attendees can address the agent, and the absent user can later watch the recording from its perspective to respond. Through two mixed-method studies with 45 participants using the Wizard-of-Oz approach, we explored both the perspectives of the attendees in the original meeting and of the absent users later re-watching the meeting. We found that the stand-in can enhance meetings, benefiting both present and absent collaborators. Present attendees can easily access information that drives decision-making in the meeting perceive high social presence of absentees. Absentees also felt included when watching recordings because of the social interactions and attention towards them. Our contributions demonstrate a proof of concept for future asynchronous meetings in which collaborators can interact conversationally more akin to how they would if it had been synchronous.


\end{abstract}

\ccsdesc[500]{Human-centered computing~Mixed / augmented reality}

\keywords{Asynchronous collaboration, Mixed Reality, Virtual Reality}

\begin{teaserfigure}
  \includegraphics[width=\textwidth]{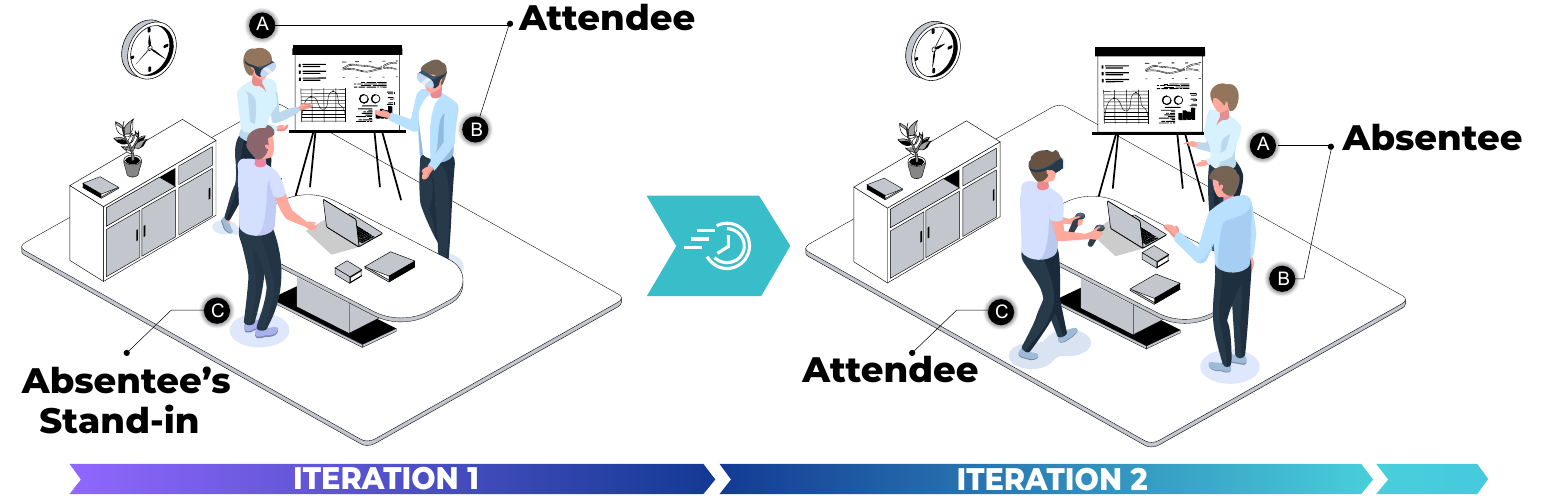}
  \caption{\rev{In a Stand-in Enhanced Asynchronous Meeting (SEAM), absent participants are represented by a virtual agent configured with their responses to the agenda items---the stand-in. Later the absentee can watch the meeting and contribute (e.g. respond and comment) from the first-person perspective of the stand-in. Those present synchronously at the meeting are referred to as \textbf{attendees}, and those watching the recording later as \textbf{absentees}. Thus, each collaborator's role depends on their attendance type per iteration. In Iteration 1, A and B attended the meeting as attendees, C could not attend due to different time zones and sent his stand-in to attend the meeting on his behalf. Later, the first meeting (Iteration 1) is finished, and C has a chance to watch the recording (Iteration 2) from the first-person perspective of his stand-in as an attendee. At this point, A and B are absent, and so are absentees. When watching the recording (Iteration 2), C can move freely and may leave comments for A and B at any time to express his thoughts or respond to unresolved questions that his stand-in could not answer in the first place.}}
  \Description{This figure illustrates two iterations of a meeting scenario. Iteration 1 shows three people: two labeled as "Attendees" (A and B) and one as "Stand-in" (C). They are gathered around a desk with a presentation board. Iteration 2 depicts a change where attendees becomes "Absentees" (A) and (B), while the former attendees now an "Absentee". (C) becomes an "Attendee". The scene includes office furniture, a clock, and a plant. A turquoise arrow between the iterations indicates progression. The layout demonstrates the dynamic nature of meeting participation, showcasing how roles can shift attendees to absentees.}
  \label{fig:SEAM-concept}
\end{teaserfigure}

\maketitle
\definecolor{note-bg}{HTML}{FFFBF0}
\definecolor{line-bg}{HTML}{FBF1CB}

\section{Introduction}

The computer-supported collaborative work (CSCW) literature has long acknowledged the importance of space and time in the organisation of collaborative activities~\cite{johansen1988groupware}. From this perspective, digital technologies have been proven critical in bringing remote users together and enabling them to share work asynchronously. A key insight from this body of work is that there is nothing inherently better about working face-to-face over remotely or synchronously over asynchronously---the mode of work depends on the nature of the task. However, the pressures of modern knowledge work have forced collaborators to work under less-than-ideal conditions: discussions that could have been quickly resolved over a face-to-face chat must often take place asynchronously over text. This has practical consequences. For example, issues like ``Zoom fatigue'' and ``nonverbal overload''~\cite{Bailenson2021Nonverbal} are consequences of the disembodied nature of remote work at scale through videoconferencing.  In this context, extended reality (XR) technologies have emerged as a way of bringing the expressiveness of face-to-face communication to remote work~\cite{vr-against-zoom-fatigue}. Though much work has focused on bringing \textit{spatially} distributed collaborators together in virtual spaces, comparatively little attention has been paid to \textit{temporally} distributed work in XR. In this paper, we propose and explore the user experience of \textbf{SEAM---Stand-in Enhanced Asynchronous Meetings}---a concept for how conversational and embodied interactions can happen asynchronously through extended~reality.

SEAM builds on recent trends in research prototypes and commercial products aimed at increasing the expressiveness of asynchronous work \cite{social-in-vr-meetings, challenges-multimodel-vr}. For example, asynchronous conversations through voice messages on WhatsApp are now the norm in many communities~\cite{ghaffary2023voice}. Loom\footnote{\url{https://www.loom.com}} supports users in recording video messages, while other collaborators can react asynchronously and leave time-stamped comments on the videos. Zoom's\footnote{\url{https://zoom.us}} AI companion can help users catch up on missed meetings by producing a conversation summary. Among research prototypes, Fender and Holz's \textit{Asynchronous Reality} explored how collaborators can leave spatial messages in mixed reality that consider causality-based changes in a mixed reality environment \cite{causality-async-reality}. Similarly, the concept of a "Ditto" explores using a personalised, embodied agent that can attend meetings on a person's behalf, representing their knowledge and appearance. It highlights the desire for more interactive and representative forms of asynchronous participation \cite{ditto}. Wang et al. then proposed ways of adapting nonverbal behaviours in recorded VR meetings to account for new participants who later watch the recording~\cite{socially-late}.  Our work brings these trends together by combining the embodied nature of XR recordings with AI-powered workflows of asynchronous video-based collaboration systems. 

We envision a SEAM unfolding through multiple iterations, with users taking the roles of \textit{attendees} and \textit{absentees}. Absentees are represented by \textit{stand-ins}, embodied conversational agents designed to represent their views in the meeting, to which attendees can address and pose questions. These stand-ins are configured with responses to agenda items that are triggered by the ongoing conversation. When later watching the recording, the initially absent user watches it from the stand-in's first-person perspective and can contribute (e.g. respond and comment) to the meeting as if they had been there. In our vision, SEAMs can happen in any recordable spatial media (XR/VR/MR) by leveraging AI-powered stand-ins, though in this paper, we focus on VR SEAMs.

This paper aims to explore questions raised by the unfamiliar features of this vision, particularly engaging in conversational interactions asynchronously, being represented by a stand-in, and interacting with someone else's stand-in in a meeting. We do so by building a SEAM prototype and conducting two mixed-methods exploratory user studies with 45 (30+15) participants to understand their experiences.  For better control of the stand-ins, we used a Wizard-of-Oz approach, manually triggering the AI behaviour. In Study 1, pairs of participants had to make a group decision while considering the perspectives of a third participant who could not attend but was represented by a stand-in. Then, participants were asked to re-watch the meeting they just had, but now from the first-person perspective of the stand-in, to elicit reflections about how they would be perceived by absent participants. In Study 2, we recruited a new set of participants who were asked to experience the meeting recorded in Study 1 as an absentee who missed the original meeting from the first-person perspective of the stand-in. These studies allowed us to explore the following research questions:
\begin{itemize}[noitemsep, topsep=0pt]
    \item RQ 1: How does the presence of the stand-in affect the social presence between attendees and the absentee?
    \item RQ 2: How do attendees interact with the stand-in? 
    \item RQ 3: How do attendees perceive the stand-in's responses?
    \item RQ 4: How do absentees perceive the attendees' social interactions in the recorded VR meeting?
    \item RQ 5: What factors make absentees feel more included in the VR meeting?
\end{itemize}

Overall, this work contributes a proof-of-concept demonstration of Stand-in Enhanced Asynchronous Meetings (SEAM), presenting both an understanding of its user experience and concrete design implications for future research. Our studies informed the development of a full functioning SEAM prototype system that enabled attendees to interact with an AI-powered stand-in in real time and later re-experience the meeting from the absentee’s perspective. Our results suggest that (1) the stand-in shows potential for enhancing the attendees' perception of the absentee's social presence thanks to its nonverbal and verbal responses, (2) attendees adapt their interaction strategies with the stand-in throughout the meeting based on their varying perceptions of its capabilities, (3) attendees perceived the stand-in's interaction as genuine input from the absentee, which facilitated decision-making process in the meeting, (4) the attendees' social interaction and attention towards the absentee made the absentee feel included when watching the recording.
Based on these results, we also discuss challenges for future deployments of SEAMs, such as trust, privacy, ethics, and granularity of asynchronous participation. Empirical insights with practical design principles and the functioning system contribute to understanding how asynchronous and embodied interactions within ubiquitous computing systems can facilitate temporally distributed collaboration.

\section{Related Work}
This work focuses on the user experience of stand-in enhanced meetings in VR. Prior work has evaluated the user experience of social interactions and collaborative tasks in terms of engagement and awareness~\cite{automated-generation-virtual-embodied-characters,deconstructing-touch,measuring-engagement-during-collaboration,through-eyes-awareness-collab-vr-platforms}. Understanding current tools that have been used in synchronous collaboration helps us to unpack the factors supporting synchronous communication and understand how they can be incorporated into asynchronous meetings.

\subsection{Non-verbal behaviours are crucial for face-to-face collaboration}
Effective communication is fundamental for successfully accomplishing collaborative goals. In addition to verbal exchanges, non-verbal cues like body language, hand gestures, facial expressions, and eye gaze play a critical role in communication. In this paper, we seek to leverage the facial expressions and other listening behaviours afforded by embodied avatars to enrich asynchronous conversational exchanges.

\subsubsection{Emotional expressions enrich social interactions with avatars in collaborative activities}
In face-to-face communication, facial expressions convey important emotional messages. For example, in emotional messages, non-verbal cues can communicate even more than verbal messages~\cite{importance-non-verbal}. Consequently, studies demonstrate that avatars capable of showing various emotions through facial expressions significantly improve emotional interactions \cite{expressive-agent}. The expressiveness afforded by embodied agents can better convey emotional messages than text. Those emotional messages are crucial for effective collaboration in collaborative virtual environments (CVEs). For example, Fabri found that an avatar's emotional expressiveness increases involvement in the interaction among CVE users and a sense of co-presence \cite{emotionally-expressive-avatars}. Further, emotional messages can also lead to richer interactions and stronger relation ties among collaborators by facilitating the interpretation of intent and meaning \cite{emotional-intelligence-affect-interpretation, emotional-intelligence-in-collaboration, social-emotional-interaction-collaborative-learning, cscl-emotional-support, emotion-understanding-during-csc}. The integration of non-verbal cues through expressive avatars in CVEs opens a new way to mimic human-like interaction more closely with the goal of improving the collaborative~experience.

Furthermore, Leong et al. explore the use of Ditto, a personalised embodied agent that serves as a visual and vocal representation of an absent participant in 2D Microsoft Teams meetings. In a Wizard-of-Oz study involving 39 collaborators across 10 different teams (preceded by six focus groups), they compared a mimetic Ditto agent, which resembled the absent person in appearance, voice, and expertise, against a non-mimetic ``Delegate'' agent. This study, limited to scenarios in which the absent individual delegates their presence to the agent, revealed that Dittos significantly improved participants' perceptions of presence, trust, and decision-making during live meetings. However, the authors did not investigate cases in which missing stakeholders would later watch meeting recordings \cite{ditto}. A key distinction from our work is that interactions in Ditto took place entirely within a 2D video-conferencing setup, whereas our work investigates embodiment and perspective switching in a 3D VR environment. The use of immersive spatial cues and the ability for absentees to later re-experience the meeting from their stand-in’s first-person perspective can change social dynamics. Additionally, a essential conceptual distinction lies in the participant's intention. While systems like Ditto focus on pure delegation, where the participant offloads their attendance entirely and SEAM is designed to support delayed participation. In SEAM, the user sends a stand-in not to replace their involvement, but to enable it at a later time, with the intention of reviewing the session and contributing more asynchronously. Attendees can perceive the absentee as more present, and absentees can feel more included and empowered to contribute asynchronously. These differences are likely to provide stronger perceptions of social presence and inclusion than in 2D synchronous-only setups.

\subsubsection{Responsive listening is an important behaviour for stand-in avatars}
In collaborative activities, responding to each other is an important interaction that helps convey each collaborator's engagement during the task; for example, Alvarado et al. found that responsive listening among peers promoted engagement by giving further explanation and reasoning in a learning-teaching scenario~\cite{responsiveness-among-peers}. Other works have also shown the importance of social responsiveness and attentiveness ~\cite{automated-generation-virtual-embodied-characters, responsive-listening-behaior}. Further, \textit{``Good listening is much more than being silent while the other person talks,''} highlighting the importance of designing avatar behaviour that would be expected from human agents, such as nodding and asking questions~\cite{what-great-listeners}.
However, this process is challenging for asynchronous collaboration systems to replicate. In these systems, present participants must address and ask questions to absent collaborators. Non-verbal cues provide essential feedback and a form of responding to the speaker without interrupting (e.g. nodding) \cite{responsive-listening-behavior}. In this work, we build upon these findings to implement listening behaviours for the stand-in avatars in order to make it more natural for meeting attendees to interact with absentees. 

\subsection{Synchronous collaboration promotes awareness of collaborative tasks}
Synchronous collaboration has many advantages, including promoting a sense of contemporaneity \cite{causality-async-reality}, reducing isolation and promoting co-presence \cite{learning-teaching-in-sync-ce}. 
In-person collaboration also allows for an easier understanding of other's actions and intentions compared to distributed set-ups. This is due to the perception of additional channels of non-verbal communication while also sharing the same physical space~\cite{grounding-in-comm, gutwin2004importance}. Prior studies into distributed collaboration have demonstrated the importance of supporting group awareness and how its presence positively impacts the underlying collaboration~\cite{sereno2022survey}. Saffo et al. proposed \textit{``through their eyes and in their shoes''} for providing group awareness across heterogeneous devices (e.g. VR platform, desktop) \cite{through-eyes-awareness-collab-vr-platforms}. Their prototype allows collaborators to step into each other's view on demand so that collaborators can maintain shared knowledge, beliefs and assumptions. Piumsomboon et al. illustrated the effects of remote avatar representations at different scales~\cite{piumsomboon2018minime, piumsomboon2019giant}. The ability for avatars to change scale ensured that their embodied behaviours (i.e., deixis) remained within the field of view of remote collaborators. 

Prior work on 'Absence Agents' has focused on mitigating short-term interruptions during synchronous XR collaboration \cite{absence-agents}, leaving open the challenge of representing fully absent participants in an asynchronous context, which SEAM addresses. In addition, Wang et al. developed an AR system, \textit{TeleAR}, that allows remote users to see a high-level view of shared resources, such as design tools and design materials \cite{mutual-awareness}. By sharing the same view and context, they found the experience created increased awareness of the shared environment. The advantages of in-person collaboration are evident in current research trends in distributed collaboration, where attempts are focused on recreating the in-person experience, such as creating full-body reconstructions of users correctly registered within shared environments~\cite{holoportation-3d-realtime, irlitti2024hybrid, pejsa2016room2room}.

Alongside each user's physical representation, the underlying communication medium also impacts the ability of collaborators to maintain real-time knowledge of each other's actions~\cite{gutwin2004importance, fussell2004gestures}. Traditionally, remote communication has predominately been delivered through text-based solutions (i.e., email, instant messaging), but they lack the support for spatial information. Recently, there has been a shift towards video-based solutions~\cite{karis2016videoconferencing, full-body-webrtc-video-conferencing}, which support the synchronous real-time delivery of each user's physical representation. However, the technology inherently limits the ability for users to convey natural nonverbal behaviours due to discrepancies between remote sites and its delivery through displays~\cite{gronbaek2021mirror, Bailenson2021Nonverbal}. As an alternative for screen-based mediated communication, embodied solutions using immersive technologies have been proposed to resolve this tension~\cite{williamson2021proxemicsVR, lee2021vr, smith2018embodied}. Proposals leverage the representation of users' faces and bodies to increase interpersonal awareness, a finding shared in studies comparing text-based and face-to-face interactions~\cite{sync-collaborative-concept-ict-awareness}.
These works highlight two important principles supported by synchronous exchanges that are missing in asynchronous ones: the importance of the body to increase the channels of interpersonal communication~\cite{grounding-in-comm, smith2018embodied} and its proxemic relationship within the interaction space~\cite{williamson2021proxemicsVR}.

\subsection{Record and replay in VR/MR is promising for supporting spatial interactions and embodied experiences.}

Unlike 2D videos, VR/MR recordings can reconstruct the environment in 3D space, providing an immersive experience of past events. VR/MR environments help participants feel like they share the same virtual space and enable richer nonverbal behaviours \cite{immersing-uni-vr}. Fender and Holz's \textit{AsyncReality} allows users to record physical events via RGB-D camera so that another collaborator can replay the events in MR, including spatial interactions involving spoken messages and instructions \cite{causality-async-reality}. They demonstrated how a user can asynchronously play back the recording, preserving the collaborator's non-verbal communication captured by the camera. The need for such a system in this scenario suggests this type of natural and intuitive asynchronous interaction with nonverbal cues is missing from typical VR/MR collaboration tools. 

Lilija et al.'s \textit{``Who put that there''} system enhances navigation and event discovery in spatial recordings \cite{who-put-there}. The system supports object trajectory tracking so users can easily view how the object was moved and repeatedly playback the specific moments when the object was changed. The study tried to solve the problem of users sometimes missing important events and finding it difficult to review what happened during the event. For example, by using this spatial recording, users can find who broke a mug or put food on the desk. Wang et al. explored how the non-verbal transformation of recorded interactions around proxemics and gaze affect the social presence of VR ~\cite{socially-late}. Their study shows that the combination of spatial accommodation and gaze enhances social presence, perceived attention, and mutual gaze, which provides insightful implications for establishing better asynchronous VR interactions. 

Compared to video recordings on a 2D screen, recordings in VR/MR can enrich spatial interactions such as pointing, grabbing, and drawing. Those interactions cannot be easily achieved in video recordings. While MR allows users to maintain awareness of their physical environment and access actual objects in the real world easily, we chose to use VR in our study as it offers a fully immersive and controlled virtual environment and reduces the complexity of adapting to different physical settings.

\subsection{The Integration of Artificial Intelligence (AI) Enables Dynamic, Context-Aware Immersive Collaboration}
AI is evolving significantly within immersive environments, advancing from merely generating content to serving as an intelligent and active partner in collaboration. This shift, known as "Generative VR" utilises AI to develop dynamic, adaptive, and context-aware VR/AR experiences that users help create, rather than consume \cite{gen-ai-meet-vr}. This change overcomes the constraints of fixed, pre-defined virtual environments by enabling AI systems to understand context, offer expert advice, and adapt content in real-time.

Shi et al., in their "CARING-AI" project, demonstrate a method of crafting context-aware AR guidelines via Generative AI (GenAI)\cite{caring-ai}. Their technique allows creators to produce animated humanoid avatars that are aware of their spatial and temporal surroundings. By walking through the area with an AR headset, the author provides the necessary context that helps the AI generate visual guidance. This demonstrates how GenAI can support remote collaboration by converting instructions from a virtual environment into context-aware guidance in a physical setting. Moreover, GenAI is facilitating the development of specialised agents that act as expert assistants or personalised tutors in immersive settings. For instance, Taylor et al. used ChatGPT-powered avatars within a VR digital-twin lab\cite{rise-of-ai-scientist}. These AI ``scientists'' are trained with specific knowledge to aid in tasks such as chemical location, equipment operation, and safety information reminder, effectively serving as virtual colleagues on demand.

As highlighted by the extensive survey from Rahimi et al., the future of immersive systems lies in this profound integration of AI as an active collaborator\cite{gen-ai-meet-vr}. Essential elements such as multimodal interaction, adaptive content, and emotionally intelligent AI will improve immersive experiences' engagement and effectiveness.

\begin{figure*}
    \centering
    \includegraphics[width=1\textwidth]{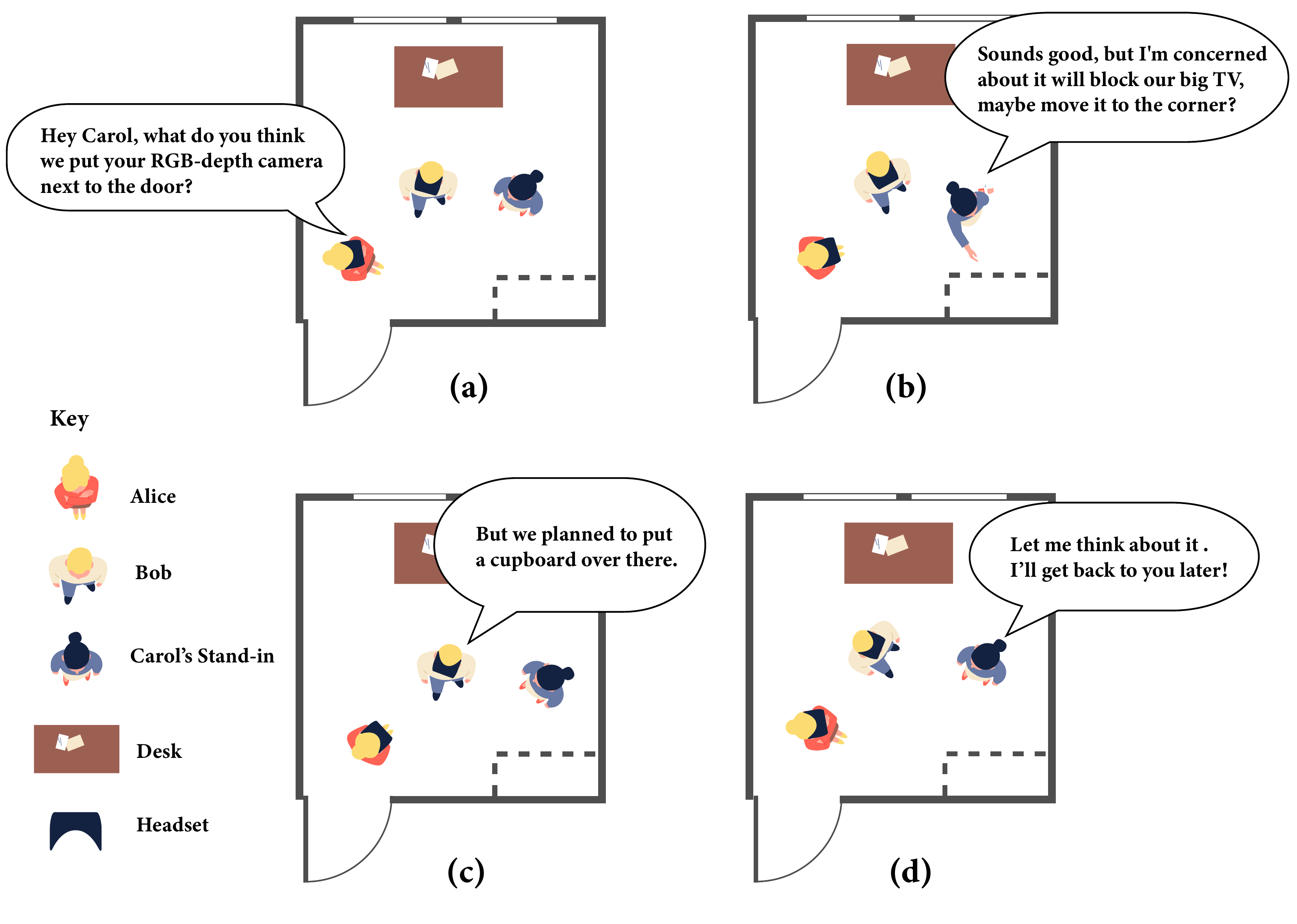}
    \caption{An example of one specific Iteration scenario in an Asynchronous Meeting: Alice and Bob attend the meeting as attendees in VR, and Carol, as an absentee, sends her AI-powered stand-in to the meeting on her behalf. The meeting topic is about their lab's interior design. (a) Alice seeks Carol's opinion about placing her equipment. (b) Carol's stand-in responds to Alice with configured responses. (c) Bob provides further information. (d) For the case that has not been prepared by Carol, Carol's stand-in suggests a later response. }
    \Description{The figure consists of four panels (a, b, c, d) from a bird's-eye view, each depicting a different stage of a conversation about arranging office equipment among three characters: Alice, Bob, and Carol's Stand-in. Each panel shows the characters in a room, represented from a top-down view, discussing the placement of an RGB-depth camera. In Panel (a), Alice suggests placing the camera next to the door. In Panel (b), Carol's Stand-in expresses concern about the camera blocking the view of a large TV and suggests moving it to the corner. In Panel (c), Bob points out that a cupboard was planned for that corner. Finally, in Panel (d), Carol's stand-in considers making adjustments later. The figure uses speech bubbles for communication and symbols indicated in a key to clarify the context of the discussion.}
    \label{fig:meeting-with-stand-in}
\end{figure*}

In summary, prior studies suggest that non-verbal cues and timely responsiveness in synchronous collaboration help collaborators promote group awareness and engagement. Recent trends highlight the benefits of VR/MR recordings in collaboration by offering an immersive environment and spatial interactions. We view the integration of embodiment capabilities from VR/MR and the intelligence of AI as promising opportunities to facilitate a seamless meeting experience in the future. Thus, we propose the concept of SEAM and explore its user experience in this study.

\section{Stand-in Enhanced Asynchronous Meetings (SEAM)}
In this section, we outline the vision for SEAMs and define their main components.  A SEAM is an asynchronous meeting that takes place in extended reality and in which participants attending asynchronously (which we refer to as the \textit{absentees}) are represented by an embodied virtual agent---their \textit{stand-ins}. To illustrate how a SEAM might unfold, consider the following use case:
\newline

\begin{mdframed}[
backgroundcolor=note-bg,
topline=false,
bottomline=false,
rightline=false,
linecolor=line-bg,
linewidth=6pt,
innerleftmargin=10pt,             
    innerrightmargin=10pt,            
    innertopmargin=10pt,              
    innerbottommargin=10pt,        
]
Three collaborators, Alice, Bob, and Carol, must discuss their new lab's interior design. Due to the spatial nature of the task, they are meeting in VR. However, Carol is travelling overseas and cannot attend the meeting due to the time zone difference. Thus, Carol sends her stand-in to attend the meeting on her behalf. Before the meeting, Carol reads the agenda and configures her AI-powered stand-in with her responses to the agenda items. Alice and Bob attend the meeting synchronously while Carol is represented by her stand-in. During the meeting, Alice and Bob asked Carol questions by directing them towards the stand-in, as shown in Figure \ref{fig:meeting-with-stand-in}. When questions related to the agenda items, the stand-in responded based on how Carol configured it. Otherwise, the stand-in suggested that Carol would respond at a future opportunity (e.g. ``I'll get back to you on that!''). Later, Carol watched the meeting from the perspective of her stand-in. Then, Alice and Bob's stand-ins replayed the conversations in the first iteration of the meeting. Carol could then confirm or amend her stand-in's automatic responses and directly reply to the questions that Alice and Bob posed. When she spoke,  the system paused the playback of the original meeting and they exhibited listening behaviours towards her instead. When Carol finished, they politely segued back into the playback. Later, Alice and Bob watched an abridged version of the first iteration, focusing on the action items and Carol's responses. The meeting proceeded back-and-forth through multiple iterations until the team decided on the final design.
\end{mdframed}

\subsection{Main components}
As the scenario illustrates, SEAMs involve participants attending the meeting synchronously and asynchronously over multiple iterations, with the presence of a stand-in. They also require new workflows, such as configuring the stand-in behaviour and watching abridged meeting recaps. In this section, we define the main components of SEAMs. We later describe how we instantiate this vision as a fully functional interactive system with the core components in Section~\ref{new-system}.

\subsubsection{Asynchronous Meetings}
We consider an asynchronous meeting to be one that takes place over multiple iterations in which one or more collaborators do not attend synchronously. An asynchronous meeting is complete when the agenda is covered and there are no further items to action. As such, all iterations of a SEAM are part of the same asynchronous~meeting.

\subsubsection{Iteration}
Unlike other types of workflow that may have different stages in collaboration (e.g. design, develop, review), we consider collaboration in an Asynchronous Meeting to be similar to collaborating on a shared Google doc \footnote{https://docs.google.com/} to which all collaborators continuously contribute. As such, a SEAM is like a living document, one that evolves through subsequent iterations. We consider an \textbf{Iteration} to be each time that the meeting takes place with at least one participant attending synchronously. We identify each iteration by an index corresponding to the sequence of contributions. After the first iteration of the meeting, each subsequent iteration builds upon the previous one. 

\subsubsection{Attendees and Absentees}
We define the roles of collaborators depending on whether they are interacting synchronously or asynchronously in any particular iteration. Those present synchronously at the meeting are referred to as \textbf{attendees}, and those watching the recording later as \textbf{absentees}. Thus, each collaborator's role depends on their attendance type in each iteration and a collaborator can be an attendee in one iteration and an absentee in another. As shown in Figure \ref{fig:SEAM-concept}, attendees in Iteration 1 become absentees in Iteration 2, and the absentee in Iteration 1 becomes an attendee in Iteration~2. 

\subsubsection{Meeting invitation}
In most calendar applications, users can respond to meeting invites by marking themselves as ``Accept'', ``Tentative'', and ``Decline'', perhaps also sending an optional message to the organiser. In a SEAM, the absentee can choose to attend asynchronously, but instead of sending a message to the organiser, the absentee can preconfigure responses to the agenda items. The stand-in can then integrate those responses into the meeting discussions as participants step through the agenda.

\subsubsection{Stand-in}
In a SEAM, each absentee is represented by a stand-in, an embodied virtual agent. The stand-in plays several roles. First, it offers a proxy for synchronous attendees to interact with the absentee and remember to include them in the decision-making process. Second, as a point-of-view from which the absentee can later watch the meeting from the first-person perspective of the stand-in. Third, as a channel for the absentee to provide their responses to agenda items. A stand-in is not a recording as its behaviours must adapt to the interaction with the attendees in real-time, with timely and appropriate responses. In its most basic form, it exhibits listening behaviours, such as looking at the current speaker, nodding, and presenting appropriate body language. However, in our vision, they can also leverage technologies like Large Language Models (LLMs) to integrate the provided responses (e.g. notes to agenda items) from the absentee into the conversational flow. Further, LLMs can help the stand-in understand the conversation and context of the meeting and respond to the questions being addressed more accurately. As Figure \ref{fig:meeting-with-stand-in} shows, Alice and Bob are attendees in this Iteration, and Carol, as an absentee, sends her stand-in to attend the meeting on her behalf and convey the ideas she prepared. In the second Iteration, Alice and Bob become absentees, and, as such, are represented by their own stand-ins. The behaviour of these stand-ins is then a combination of the recording playback and adaptations to the new context (e.g. pausing the playback when interrupted and showing listening behaviours when the attendees speak).

\subsubsection{Abridged review}
\begin{figure}
    \centering
    \includegraphics[width=0.7\linewidth]{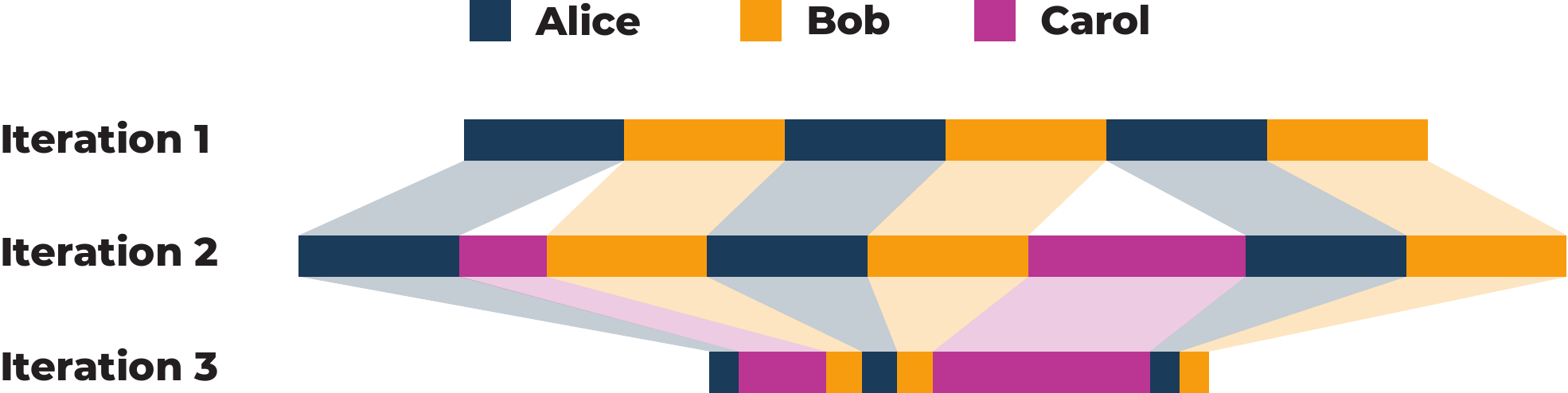}
    \caption{Example of a SEAM timeline. In Iteration 1, Alice and Bob meet with Carol's stand-in. In Iteration 2, Carol watches the recording and provides her input. When Alice and Bob revisit the meeting in Iteration 3, they only have to re-watch an abridged summary of the first iteration, but Carol's contributions in their entirety. }
    \Description{This figure illustrates a SEAM (Sequential Asynchronous Meeting) timeline across three iterations. It uses a horizontal bar chart format with different colours representing Alice (dark blue), Bob (orange), and Carol (pink). In Iteration 1, the timeline shows alternating segments of Alice and Bob, indicating their meeting without Carol. Iteration 2 expands to include Carol's input, shown as pink segments interspersed with Alice and Bob's segments. Iteration 3 is condensed, showing shorter segments for all three participants, representing an abridged review of the previous iterations with Carol's full contributions included.}
    \label{fig:SEAM-timeline}
\end{figure}

As the iterations of SEAMs progress and more contributions are incorporated, they also become longer and longer. In addition, it is difficult to envision participants wanting to re-watch meetings that they attended in the first place. To address these limitations, we envision SEAMs treating time as design material, leveraging LLMs to create abridged versions of meetings that fit into the time they have available, compressing or expanding their content accordingly, as shown in Figure~\ref{fig:SEAM-timeline}. This way, users can interact as they would with a media player, easily navigating to new content and viewing specific agenda items accompanied by text-based summaries generated by LLMs. Additionally, in the vision of SEAM, it should support the generation of non-verbal behaviours for the stand-in during abridged review. Using LLM-derived summaries as input, the system synchronises synthesised speech with simple head nods, gaze shifts, and gestural cues to reflect emphasis and engagement. While the non-verbal animations are currently limited in variety, they provide an embodied sense of presence that supports engagement during playback. Future work could incorporate generative animation to produce more contextually appropriate behaviours, helping the stand-in feel more lifelike and expressive during summary delivery \cite{real-time-animation, make-an-animation}. 

In summary, the aforementioned main components contribute to the vision of SEAM. We instantiate this vision as two systems. First, we built a simpler version as a technology probe for the user studies, which we describe in Section 4. We then incorporated the learnings from this study into a full system, described in Section 9.
The technology probe did not incorporate LLM or AI in the stand-in, choosing instead to control them manually to minimise variability in performance during the studies. In contrast, the full system includes an LLM-powered agent. The studies presented in the next sections focus on the experience of the absentees, attendees, and their interactions with and through the stand-in. We do not study the secondary features of the approach in this paper, such as the meeting invitation, agenda configuration, and abridged review, leaving these for future work. 

\section{Technology Probe Implementation}
We instantiated our vision of SEAM through a functional technology probe, focusing on a two-iteration SEAM, where an original meeting is recorded and later watched by an absentee who had been represented by a stand-in.

The VR prototype was developed in Unity 2022.3 LTS, running on an off-the-shelf AORUS Gaming laptop (Windows 11, Intel Core i7-12700H, 32GB RAM, NVIDIA GeForce RTX 3080Ti GPU) and deployed to a Meta Quest Pro with face, eye, and hand tracking enabled.  Each VR instance is connected through the Photon Unity Networking (PUN) Fusion 1 framework, which supports the synchronization of each user's movements (hand gestures, facial expressions, etc) and Photo Voice to stream the voice in the meeting. In addition to supporting all interactive components of the system, one Windows machine running the developed Unity project is also responsible for recording the meeting, a process that captures all attendees' embodied and verbal communication within the virtual environment. 

\subsection{Multi-user VR meetings}
We used the Meta Movement SDK (ver. 57) to enable face, eye, and hand tracking, which allows participants to have natural facial expressions and body movements. Meta Avatars (ver. 20.3) SDK is used to represent users' actions in the system. The prototype was deployed on a Meta Quest Pro head-mounted display. To increase realism, all movements (face, eye, and hand) tracking were enabled during both user studies. For supporting real-time multi-user remote meetings, we used the Photon Unity Networking (PUN) Fusion 1 framework to synchronise users' movements and Photon Voice to stream the voice in the meeting. Participant's movements were all captured by the headset, and their voice was captured by the headset's microphone. PUN's Shared Mode was used to create the meeting room in Photon's cloud, therefore there is no specific host role in the system. The motion data is derived from Meta Avatar's API as a byte array by frame and this data is broadcasted to other clients. Once each client receives the movement data, the data is applied to the avatar's instance accordingly. This process is running per frame to stream users' actions through networking. Using PUN's latency log API, the average Round Trip Time (RTT) from the user side is 100 ms.
\subsection{VR meeting recordings}
The recording process is run on the Windows computer, instead of the VR headset. We integrated the recording feature in the Unity project's editor. Thus, to record the meeting, a Windows computer must initiate the meeting as an invisible user by playing the Unity project in Editor. When the Windows computer joined the meeting, the researcher manually removed the T-pose avatar (as there was no tracking for this avatar) and spawned the stand-in in the meeting. The initial looking-around animation of the stand-in was automatically started upon its spawn. As this Windows computer can be considered as a user without using the VR headset, it can also receive audio and animations from other users through networking. Thus, we developed functions to record the audio (sample rate 48 kHz) and animations (72 FPS) separately. To make sure the recording frame rate matches the playback frame rate, we manually set the target frame rate (72 FPS) for both scenes (recording and playback) in the Unity project. The audio file stores users' speech in the WAV format. The animation file stores users' full-body actions and movements as a JSON file. For both user studies, we used the PC for recording the meetings. 
\subsection{Playback VR recordings}
We also integrated the playback feature in the Unity editor. As the recording files were stored on the Windows computer, the playback process was performed on the computer instead of the VR headset. To view the recordings in VR, the headset was connected to the Windows computer via Quest Link. Once the participants got into the dashboard of Quest Link, the researcher played the Unity project and participants saw the scene from the same position that the stand-in stood in Iteration 1. In Study 1, we set up the same Unity project on two Windows computers, one gaming laptop and one PC. For Iteration 1, we only used the PC to record the meetings. After Iteration 1, the researcher shared the meeting recordings from the PC to the laptop, so those two Windows computers had the same copy of the recording files. In Iteration 2, each participant's headset was connected to one Windows computer, then the researcher played back the recording for each participant. As the playing back function was running locally on the computer, each participant watched the same meeting content independently.

\subsection{User workflow for attendees and absentees}
The user experience of the prototype is slightly different depending on whether they are an Attendee or Absentee in Iteration 1.
For Attendees, once they load up the VR prototype, they can select an avatar and join the meeting, interacting and communicating naturally using embodied gestures with their peers. Once the meeting is complete, their recorded conversations will be available for review by any participants.

\begin{figure*}
    \centering
    \includegraphics[width=1\linewidth]{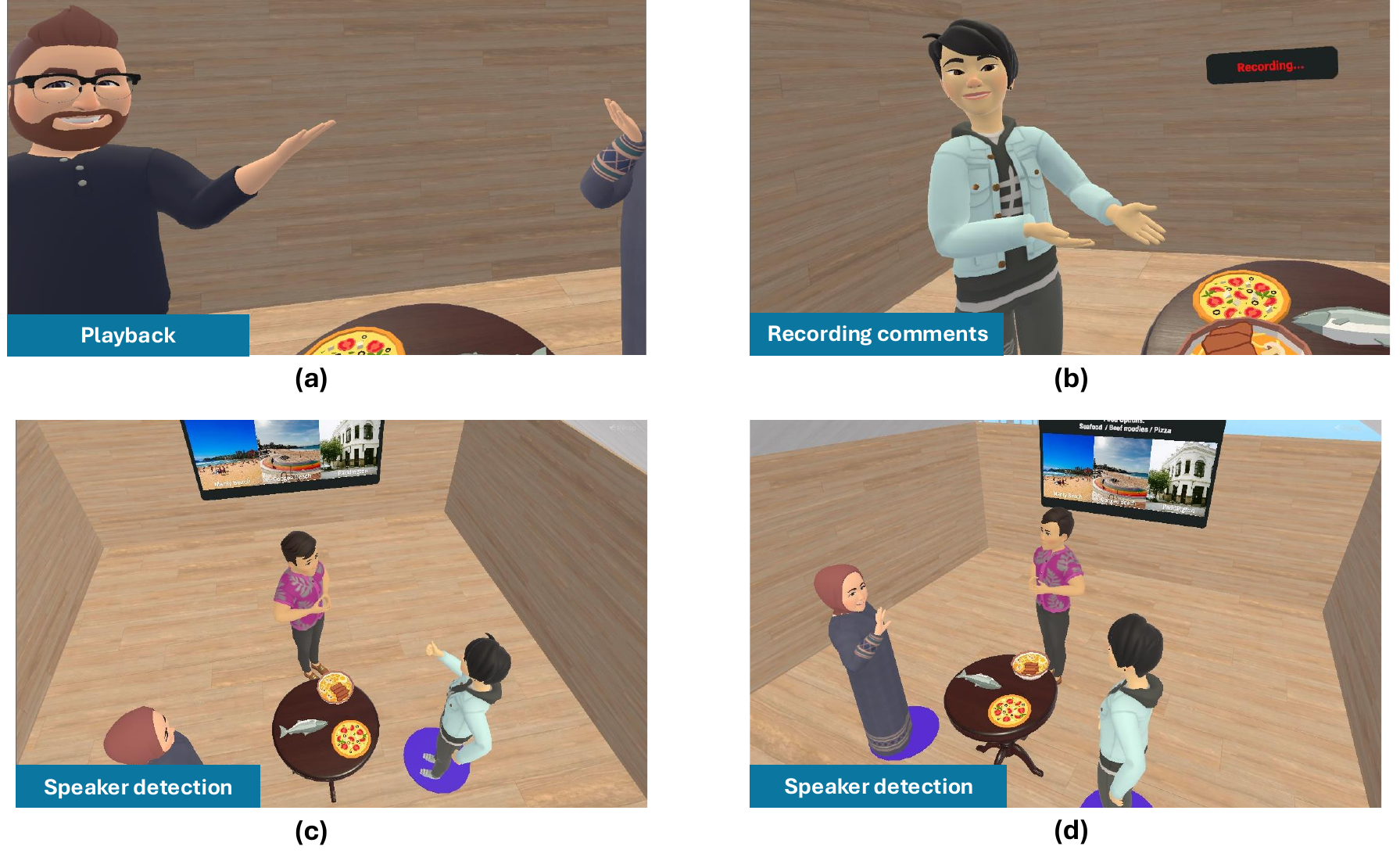}
    \caption{System features: (a) When the system plays back the recorded meeting, users can see the scene from the first-person perspective of the stand-in at the same position it stood. (b) The recording indicator is displayed when the system is recording the user's comments, including actions and speech, during the recorded meeting. (c) and (d) Stand-in detects the active speaker and turns to face the speaker with listening behaviour   (e.g. nodding). }
    \Description{This figure displays four different scenarios depicted through animated avatars in a virtual environment, illustrating various system features for a SEAM. Image (a) shows an avatar (male) replaying a recorded meeting from an individual perspective, as indicated by the label "Playback." In image (b), a female avatar is shown speaking with a "Recording" indicator above her head, illustrating the feature of recording comments during a meeting. Images (c) and (d) both depict scenes with multiple avatars standing around a table with pizza, and stand-in looks at the active speaker, labelled "Speaker Detection".}
    \label{fig:system-features}
\end{figure*}

\begin{figure}
    \centering
    \includegraphics[width=1\linewidth]{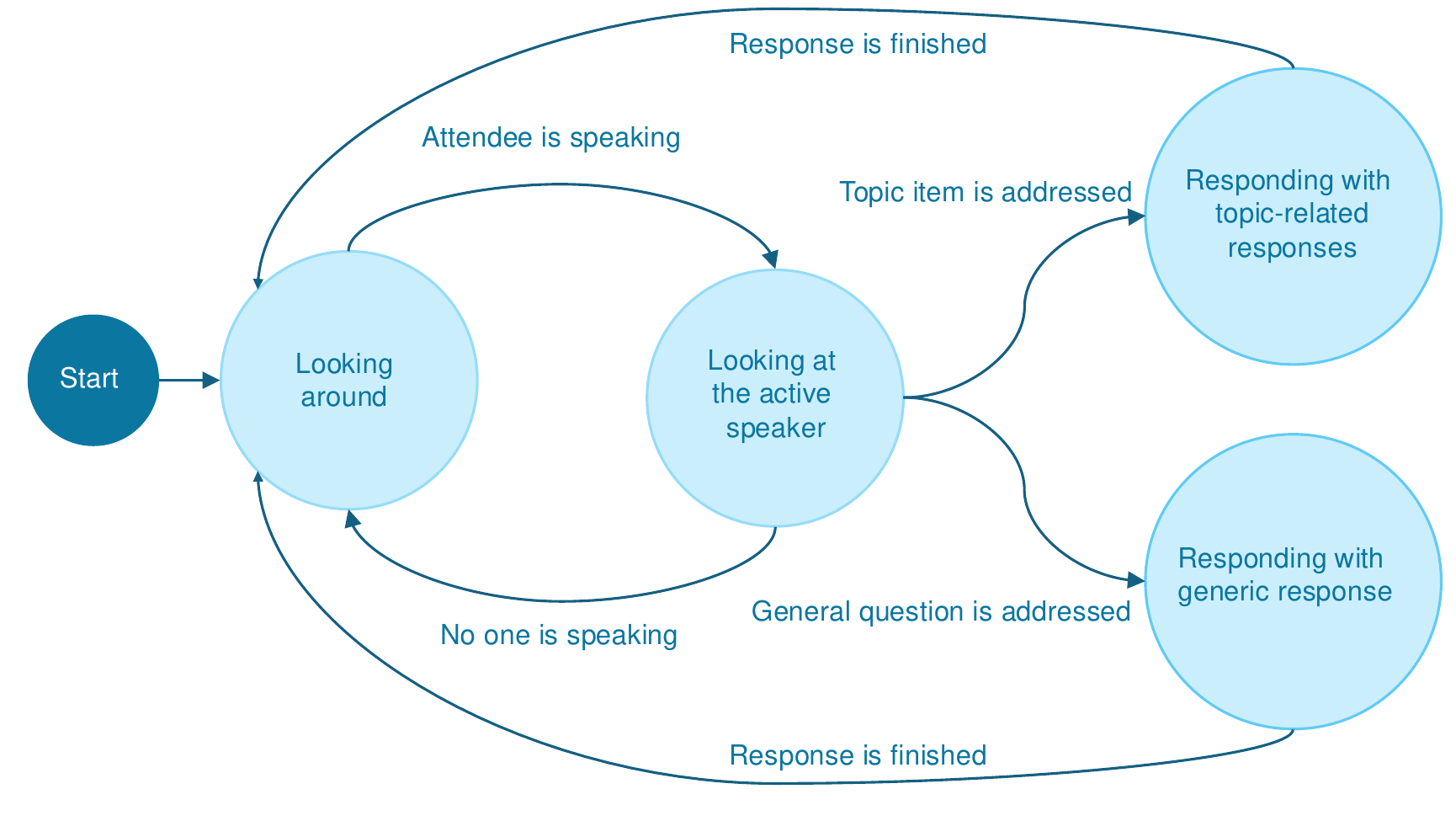}
    \caption{Stand-in's state machine: when stand-in is initialized, it starts looking around as idle movements. When an attendee is speaking, the stand-in will turn to look at the active speaker while nodding. If no one is speaking, the stand-in goes back to the ``looking around'' state. If an attendee asks stand-in questions related to the meeting topic's item that has been pre-configured, the stand-in will respond to this attendee with non-verbal and verbal cues. For other general questions, stand-in will just suggest a later response. Whenever the response is finished, the stand-in will return to the Looking around state.}
    \Description{This figure shows a state machine diagram for the "stand-in". It illustrates the behaviour flow of the stand-in, starting from an idle "Looking around" state. When an attendee speaks, it transitions to the  "Looking at the active speaker" state. If no one is speaking, it returns to looking around. From the active speaker state, it can respond in two ways: with topic-related responses if a relevant topic is addressed or with a generic response for general questions. The diagram uses circular nodes for states and arrows to show transitions, with labels explaining the conditions for each transition.}
    \label{fig:state-machine}
\end{figure}

For Absentees, their selected stand-in representation joins the meeting on their behalf. During the meeting, the absentee is not involved; the stand-in is responsible for responding naturally to prompts from other Attendees. Such responses include updating body posture based on the detection of active speakers, as shown in Figure \ref{fig:system-features}c and Figure \ref{fig:system-features}d, and providing suitable verbal responses to meeting criteria based on provided discussion points. To facilitate responsive body movements, a state machine indicates how the stand-in should behave during the meeting using prerecorded idle and response movements (Figure \ref{fig:state-machine}). Verbal responses are generated using PlayHT \footnote{https://play.ht}, an AI voice generator platform.

At a later moment in time, the Absentee can join the recorded meeting asynchronously, where they experience the meeting from the first-person perspective of their stand-in (in this phase, they are now an Attendee). During this experience, they have the ability to press a button on their Quest controller to pause the recording (see Figure \ref{fig:system-features}b), and capture their response including their full-body actions and speech in recognition to the current topic within the meeting. The response is timestamped to ensure it is included in the larger meeting at the right moment.


\begin{figure*}
    \centering
    \includegraphics[width=1\linewidth]{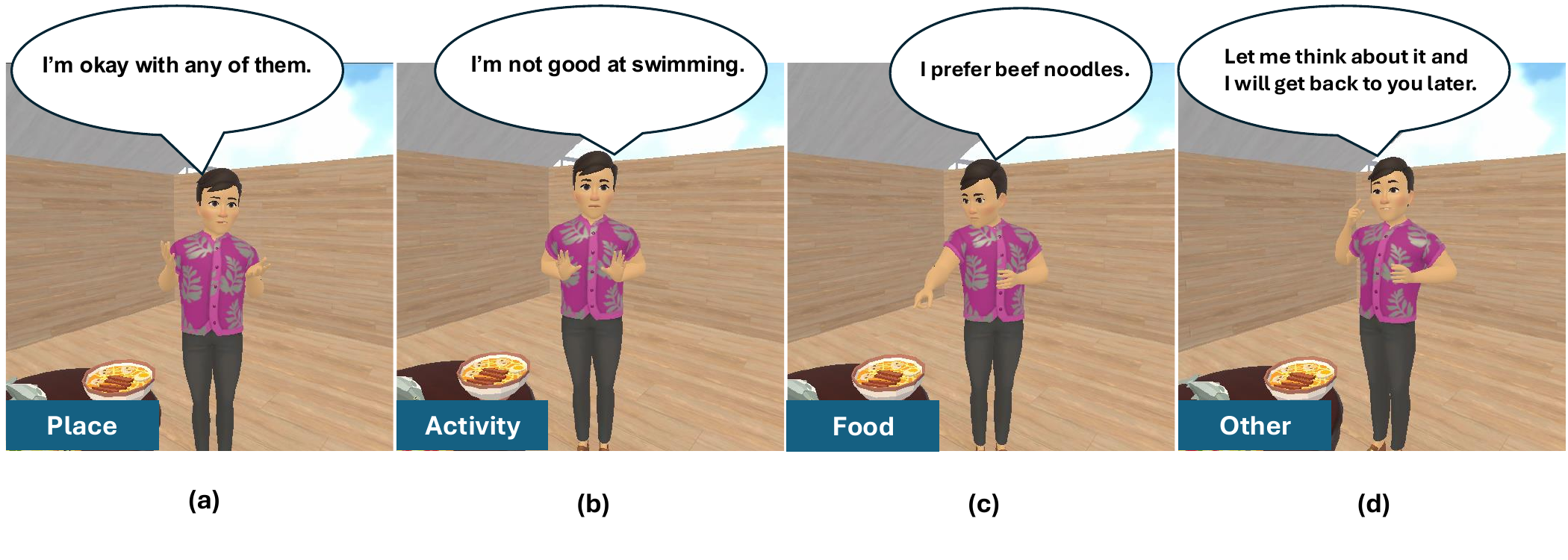}
    \caption{Stand-in's responses: if the stand-in is addressed regarding (a)  Place, the stand-in will respond while shrugging; (b) Activity, the stand-in will respond while waving his hands, (c) Food, the stand-in will respond while pointing at the beef noodles model, (d) Other cases, the stand-in will respond while pointing head and suggest a later response. }
    \Description{The image depicts a figure, a stand-in character in a virtual reality environment, demonstrating different responses based on categories like Place, Activity, Food, and Other. The figure appears in (a) it responds to a question about Place by shrugging and saying, "I'm okay with any of them." In (b), addressing Activity, the stand-in gestures with a hand wave, stating, "I’m not good at swimming." In (c), responding to Food, the stand-in points to a model of beef noodles and comments, "I prefer beef noodles." Finally, in (d), when addressed with Other queries, the stand-in points to its head, indicating thought, and says, "Let me think about it, and I will get back to you later." Each panel corresponds to various human-like gestures and communication modes depicted in a simplified 3D environment.}
    \label{fig:stand-in-responses}
\end{figure*}

\section{Study 1 - The experience of the attendees}
This paper presents the results of two studies; their methods are first described in the following two sections, while a combined discussion of their results is provided in Section \ref{sec:results}.

To explore the user experience of having a meeting with a stand-in, we conducted an initial exploratory laboratory user study. Study 1 included two tasks that required participants to attend an asynchronous meeting over two iterations. In the first task, pairs of participants were asked to have a meeting with a stand-in. In the second task, participants watched the meeting they just had from the stand-in's first-person perspective. This task gave participants an understanding of how they would later be perceived by the absentee in order to elicit self-reflection about their behaviours. We chose not to frame this study as a controlled experiment with a baseline condition because our goal was exploratory: to understand how people would respond to the novel concept of a stand-in agent in asynchronous meetings. Rather than isolating specific variables for measurement, we treated SEAM as a \textit{technology probe}, a deliberately provocative prototype designed to surface unexpected behaviours, reactions, and design opportunities. This approach allowed us to explore how participants interpreted the presence and actions of the stand-in, how it shaped their sense of inclusion and communication, and what tensions emerged. A controlled comparison at this early stage would have constrained the space of inquiry and overlooked the kinds of rich, qualitative insights that are most useful when investigating an unfamiliar interaction paradigm.

The design of these two tasks helped us seek the answers to how attendees perceived the absentee's social presence when it was represented by a stand-in (RQ1), how attendees would interact with the stand-in (RQ2), and how attendees perceived the stand-in's response when it was being addressed (RQ3). 

To control for the variation of behaviour and triggering errors produced by AI systems, we opted to use a Wizard-of-Oz approach to manually trigger the stand-in's contributions. The first author monitored the conversations during the meeting and manually played back recorded responses (see Figure \ref{fig:stand-in-responses}). This decision better aligned with our motivations in understanding users' reactions to the general concept of SEAM. For verbal responses by the stand-in, we used a male character, Lachlan, who speaks English.

Throughout the study, participants and the researcher were in the same room, and the entire study lasted approximately 90 minutes. The study received ethics approval from our university. 

\subsection{Participants}
We recruited 30 participants in pairs to undertake the study. The participants were aged between 19 and 39 years old, with a mean age of 23 (SD=4). Among them, 15 identified as men and 15 as women. Participants registered for the study in pairs so all participants in the same session knew each other before undertaking the study. Nineteen participants had MR/VR experience. Participants received a \$30 AUD voucher for their time.

\subsection{Method}
\begin{figure*}
    \centering
    \includegraphics[width=0.7\textwidth]{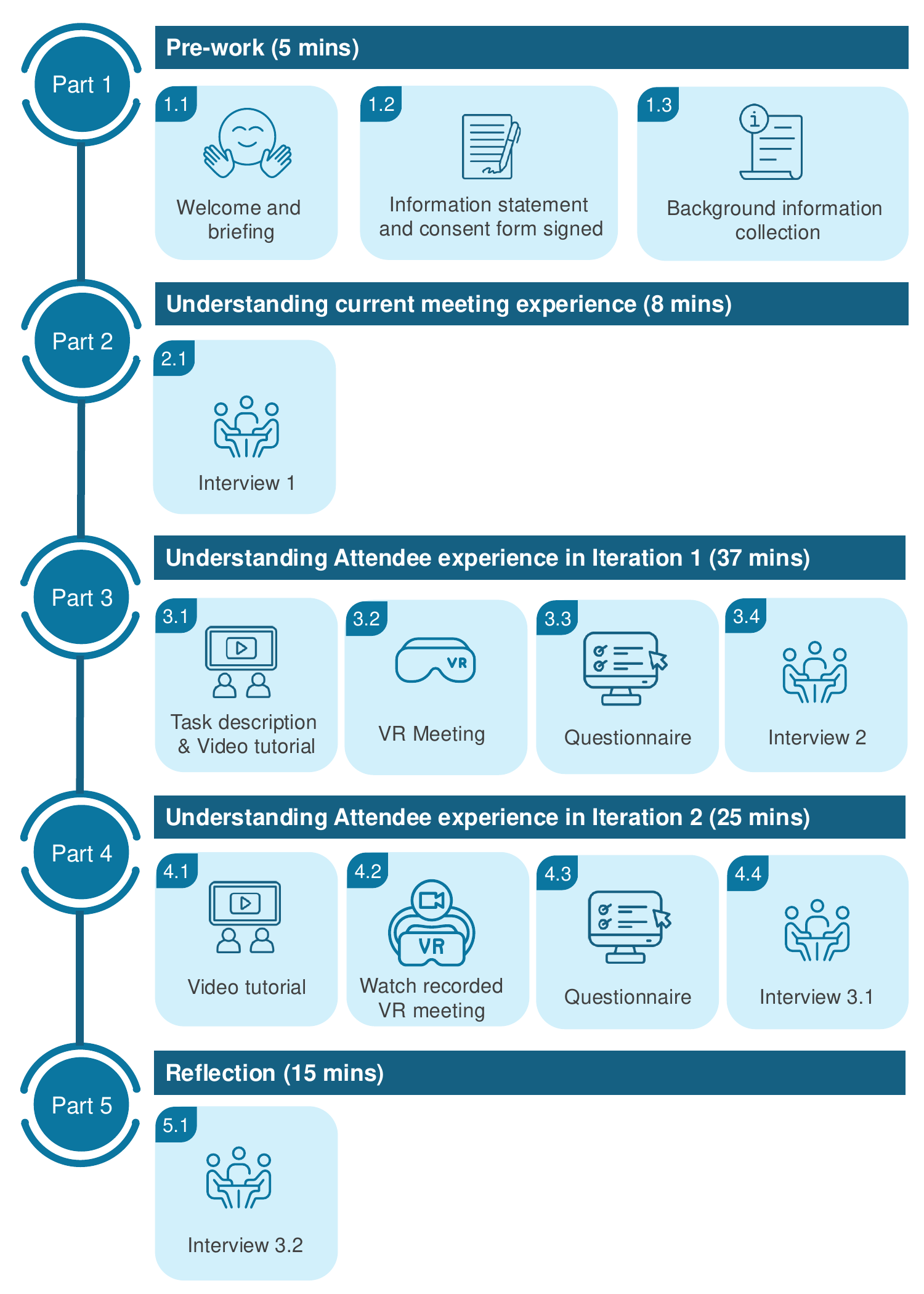}
    \caption{A visual summary of the Study 1 flow}
    \Description{This image represents a visual summary of the user study flow, structured into five main parts, each designed to assess different meeting experiences, particularly involving SEAM. Part 1, termed "Pre-work," involves welcoming and briefing the participants, getting consent forms signed, and collecting background information, all within five minutes. Part 2 focuses on understanding participants' current meeting experiences through a brief interview lasting eight minutes. Part 3, which takes the longest time at 37 minutes, dives deeper into attendees' experiences in VR meetings, including a task description, a VR meeting, a questionnaire, and a second interview. Part 4, lasting 25 minutes, continues the exploration into attendee experiences in Iteration 2 with a viewing of a recorded VR meeting, another questionnaire, and a third interview. The study concludes with Part 5, a 15-minute reflection period involving a final interview. The image uses icons and a timeline format to provide clarity and structure to the study's flow.}
    \label{fig:user-study-flow}
\end{figure*}

The study included five parts, as shown in Figure \ref{fig:user-study-flow}.

\subsubsection{Part 1: Pre-work} \label{pre-work}
Upon their arrival, the researcher welcomed the participants, introduced the study, provided a plain language statement, and obtained written consent. Then, participants were asked to fill out a background information~form.

\subsubsection{Part 2: Current meeting experiences} \label{current-meeting}
The researcher started the first interview with two participants (questions are listed in Table \ref{tab:first-interview} in the Appendix). At the beginning of the interview, the researcher asked participants to recall a meeting in which at least one member was absent and use this meeting's experience to answer the interview questions. 

\subsubsection{Part 3: Attendee experience in Iteration 1} \label{part3}
Participants were provided the context of the first exercise: the meeting was about planning a weekend with a fictional friend, Lee, who was not able to attend and sent his stand-in instead.

Each participant was provided with two printed sheets. The first sheet was the same for both participants, containing information about options for place, activity, and food. The second sheet provided information specific to each participant about their own preferences, which they should advocate for through negotiation. The researcher explained to participants that they could address Lee's stand-in during the meeting and that by the end of the meeting, they should have agreed upon a place, activity, and food, taking all three participants' preferences into account. Participants were also informed that if they needed a break or faced technical issues, they could raise their hands, and the researcher would assist them. All information sheets can be found in the supplementary materials. 

\begin{figure*}
    \centering
    \includegraphics[width=1\textwidth]{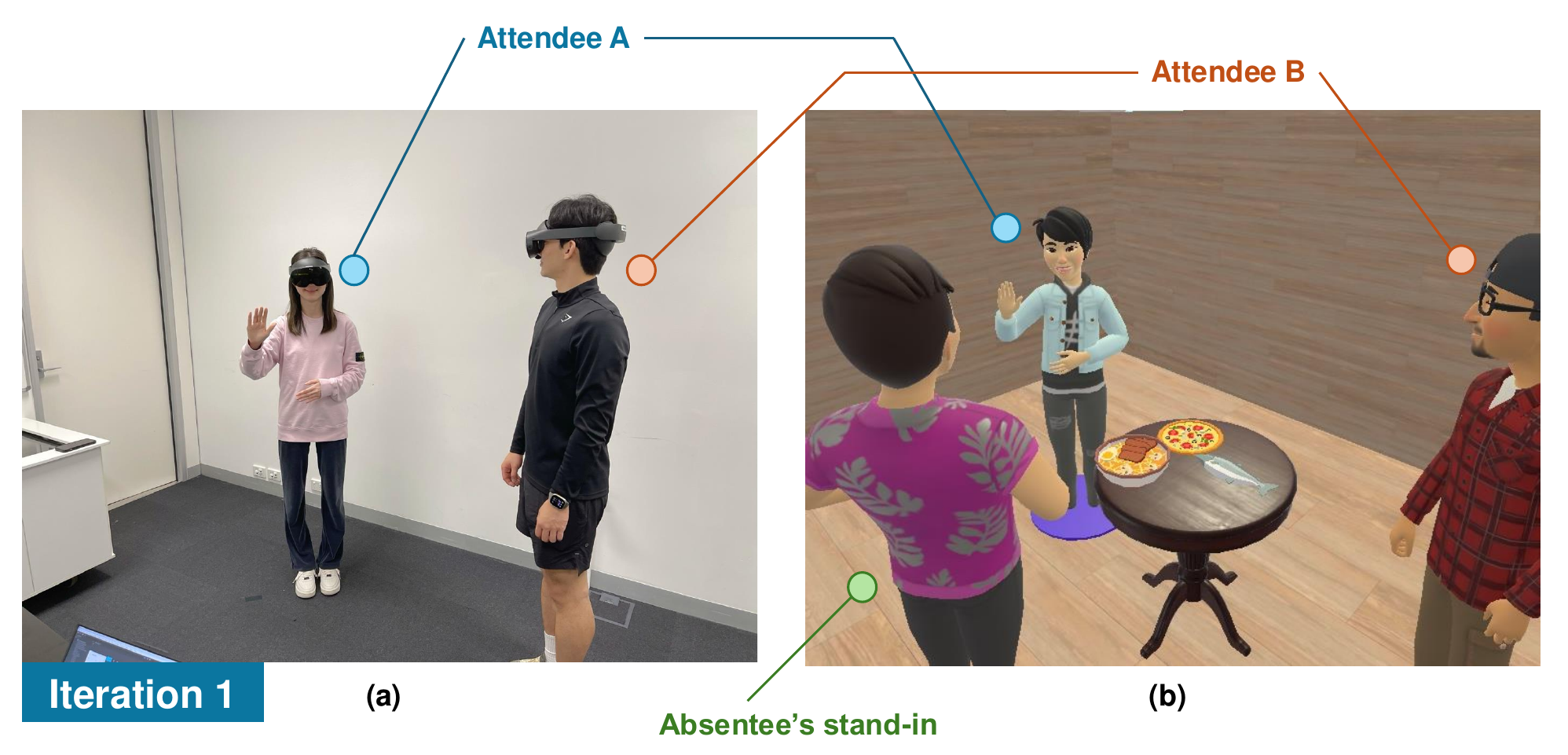}
    \caption{User study scenario for Iteration 1: (a) Two participants attended the meeting as attendees. Lee, their fictional friend, sent his stand-in to attend the meeting on his behalf. (b) Attendees were interacting with Lee's stand-in by saying ``Hi!''. }
    \Description{This figure illustrates a user study scenario for Iteration 1 of a SEAM. It consists of two images labelled (a) and (b). Image (a) shows two people wearing VR headsets in a physical room, representing Attendee A and Attendee B. Image (b) depicts a virtual meeting environment with three avatar figures. One avatar, standing in the centre, represents the stand-in for an absentee named Lee. Two other avatars, representing Attendees A and B, are shown interacting with Lee's stand-in. The virtual room includes a small table with what appears to be food items.}
    \label{fig:study-setup-1}
\end{figure*}
After the task introduction, participants watched a short tutorial showing basic hand gestures for opening and closing the application and a demonstration of how to wear the headset. Then, the researcher assisted them in wearing the headsets and initiated the study, as shown in Figure \ref{fig:study-setup-1}. During the meeting, the researcher did not intervene in the discussion, and the recording was stopped when participants confirmed their decision. The VR meeting lasted approximately 7 minutes. Then, participants were asked to fill out the Networked Minds \cite{networked-minds} and a self-designed questionnaire (Table~\ref{tab:self-designed-questions}). In the second interview (Table \ref{tab:second-interview}), the researcher asked about the experience of interacting with the stand-in and how it compared to other cases in which a required participant was absent.

\subsubsection{Part 4: Attendee experience in Iteration 2}
\begin{figure*}[h]
    \centering
    \includegraphics[width=1\linewidth]{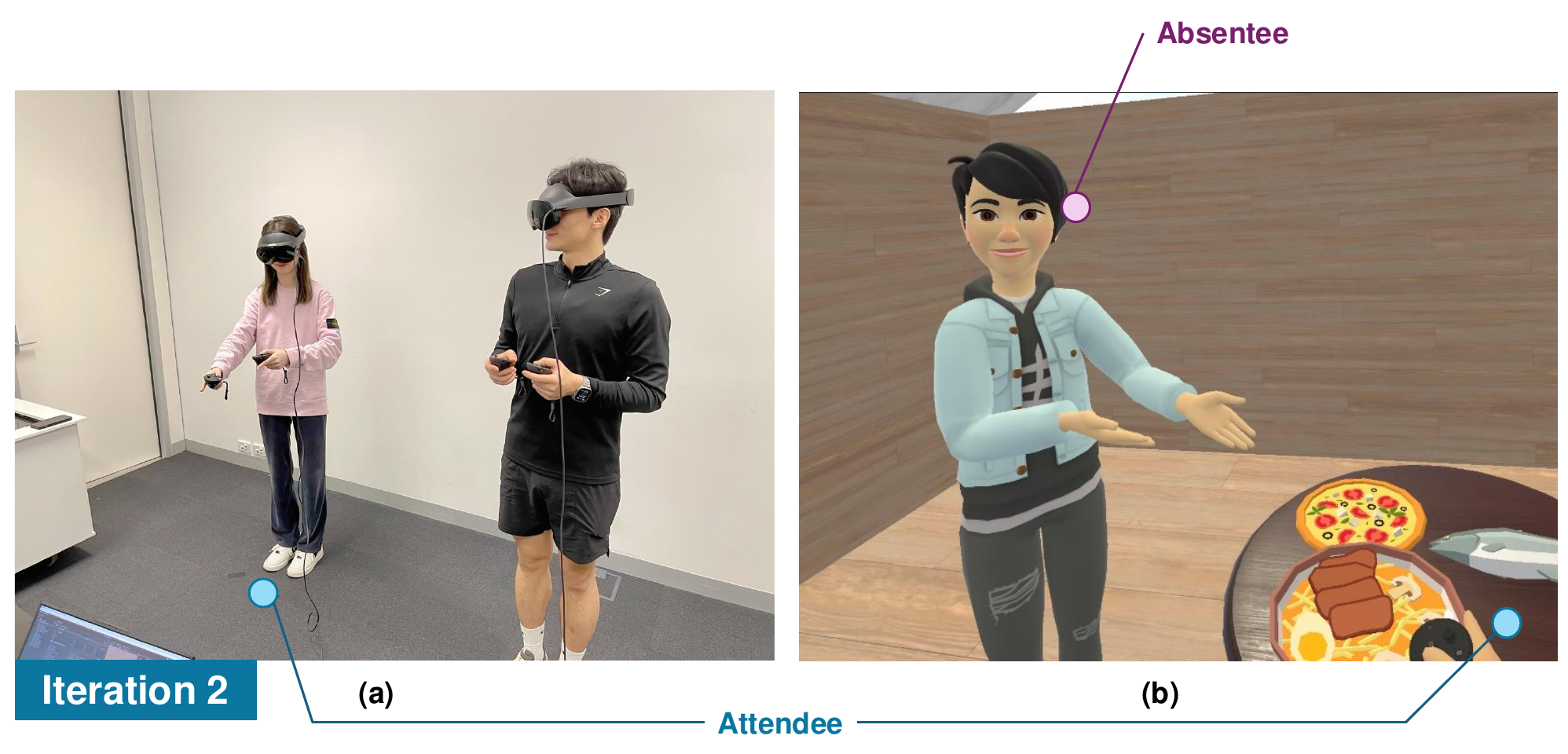}
    \caption{User study scenario for Iteration 2: (a) Participants watched the recorded meeting they just had independently but in parallel. In Iteration 2, participants watched the recording from the stand-in's perspective. (b) First-person perspective from the participant's point of view when watching the recorded meeting.}
    \Description{This figure illustrates a user study scenario for Iteration 2 of a SEAM. It consists of two images labeled (a) and (b). Image (a) shows two people wearing VR headsets in a physical room, representing participants watching a recorded meeting separately but in parallel. Image (b) depicts a first-person perspective of the VR environment, showing an avatar representing the absentee participant. The avatar is gesturing near a table with virtual food items, including pizza. The figure demonstrates how participants can review a recorded VR meeting from stand-in's perspectives.}
    \label{fig:study-setup-2}
\end{figure*}

Participants were shown how to pause, record, and resume the recording using a controller. Then, participants were again provided with a Meta Quest Pro and controllers to perform the second VR activity independently
(Figure \ref{fig:study-setup-2}).
In the second VR activity, participants were asked to experience the meeting they just had through the eyes of Lee, the stand-in. They were also allowed to record responses or comments if they desired. Each participant's headset was connected to a Windows computer via a Quest Link cable running an independent instance of the prior meeting. Thus, while the participants watched the same meeting content, their interactions and control (e.g., pausing or resuming) of the recording would not affect the other participant's experience.

At the conclusion of the meeting, participants were asked to remove their headsets and fill out the last self-designed questionnaire (Table~\ref{tab:self-designed-questions}). In the third interview, the researcher asked about their experience of watching the meeting and how they felt when being addressed (Table \ref{tab:third-interview}).

\subsubsection{Part 5: Reflection}
In this last interview, questions related to the overall experience as shown in Table \ref{tab:third-interview}. This interview concluded with the researcher answering participants' questions about the study.

\section{Study 2 - The experience of the absentees}
In Study 1, participants watched the meeting they just had from a stand-in's perspective to help them visualise how the absentee would perceive their actions. However, given that these participants had just attended the meeting, this experience might differ from the one of a true absentee. Therefore, we conducted a second study for participants not involved in the original meeting to watch the meetings recorded in Study 1 as absentees. Study 2 was designed to understand how an absentee would perceive the social interactions that occurred in the VR recordings (RQ4) and what factors made them feel included when watching the recordings (RQ5). In addition to validating the findings of Study 1, it helped us better understand the factors that affect perceptions of inclusion and exclusion. The study lasted approximately 60 minutes and received ethics approval from our university.

\subsection{Participants}
We recruited 15 participants who did not participate in Study 1. The participants were aged between 19 and 40 years old, with a mean age of 26 (SD=5). Among them, 7 identified as men and 8 as women. Eleven participants had MR/VR experience. Participants received a \$30 AUD voucher for their time.
\subsection{Method}
In Study 1, we recorded 15 groups' meetings, and those meetings were used in Study 2 for watching. Study 2 had a similar structure as Study 1. 
\subsubsection{Part 1: Pre-work \emph{\&} Part 2: Current meeting experiences}  We followed the same procedure used in Study 1 (Sections~\ref{pre-work} and~\ref{current-meeting}).

\subsubsection{Part 3: Attendee experience in Iteration 2} \label{study2-part-3}
\begin{figure}
    \centering
    \includegraphics[width=1\linewidth]{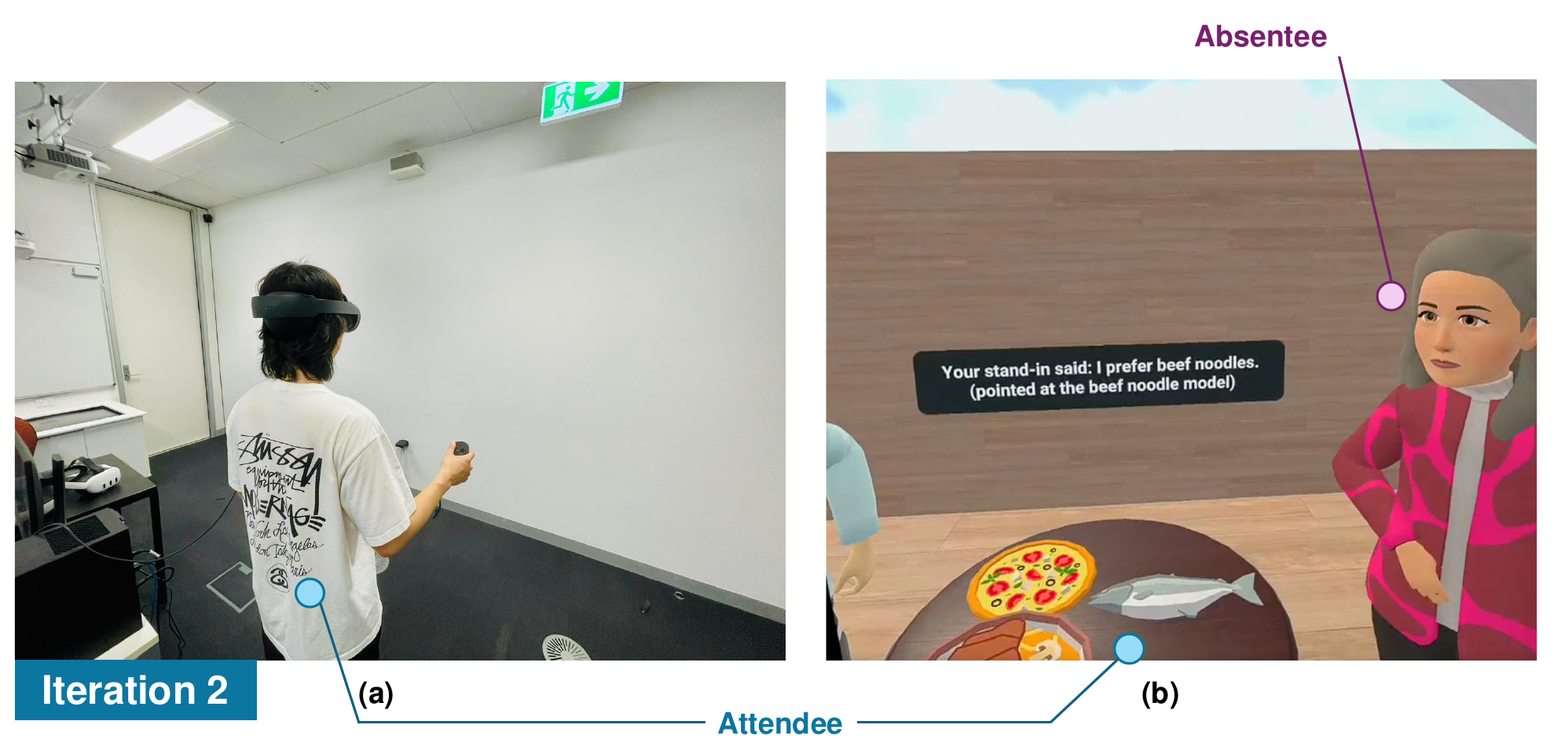}
    \caption{(a) Study 2 setup that participants watched the meeting recorded in Study 1. (b) View from the participant's first-person perspective and an example of how the stand-in response was displayed in a text panel for the attendee in Iteration 2 while watching the recording.}
    \Description{This figure shows a person using Meta Quest controllers and wearing Meta Quest Pro to perform the Study 2 tasks. The second image shows the view from the View from the participant's first-person perspective.}
    \label{fig:stand-in-res-text}
\end{figure}

The context for the task was that the participant had sent a stand-in to a weekend planning meeting they could not attend on their behalf, and they now needed to watch the recording as an absentee. The researcher provided what the stand-in received about the agenda items in Iteration 1 (see Figure \ref{fig:stand-in-responses}). Participants were shown how to pause, record, and resume the recording using a controller. Then, participants were provided with a Meta Quest Pro and controllers, and they were instructed that they could record responses or comments if they desired. The participant's headset was connected to a Windows PC via a Quest Link cable. Participants were told they would experience two different meetings, which occurred as part of Study 1. While watching each recording, participants were able to see their stand-in's responses in a text panel, as shown in Figure~\ref{fig:stand-in-res-text}. After each meeting, participants were asked to fill out a Networked Minds questionnaire. Upon finishing the questionnaire after the second meeting, Study 2 ended with the final interview about the experience of watching the recordings (Table~\ref{tab:study-2-interview-2}). 

\section{Results}
\label{sec:results}
We collected quantitative and qualitative data in both studies. As quantitative data, we measured social presence with Biocca and Harms's \textit{Networked Minds} \cite{networked-minds} questionnaire. In \hyperref[part3]{Study 1 Part 3}, participants completed the Networked Minds questionnaire twice with different agent terms. In the first version, the agent term in each question was the other attendee's name; in the second version, the agent term was the absentee's name, \textit{Lee}. In \hyperref[study2-part-3]{Study 2 Part 3}, participants completed the questionnaire with the agent term, "the other two participants", referring to the absentees in the recordings (Iteration 2). Our qualitative data included interview and written observation data from participants' VR tasks. We analysed it using a general inductive approach~\cite{general-inductive-approach}, re-reading transcripts, refining the categories of codes, watching the recordings, and identifying key themes relevant to the research questions and SEAM vision.

Our analysis was guided by the following research questions:

\begin{itemize}[noitemsep, topsep=0pt]
    \item RQ 1: How does the presence of the stand-in affect the social presence between attendees and the absentee?
    \item RQ 2: How do attendees interact with the stand-in? 
    \item RQ 3: How do attendees perceive the stand-in's responses?
    \item RQ 4: How do absentees perceive the attendees' social interactions in the recorded VR meeting?
    \item RQ 5: What factors could make the original absentee feel more included in the VR meeting?
\end{itemize}

\begin{figure}
    \centering
    \includegraphics[width=0.5\linewidth]{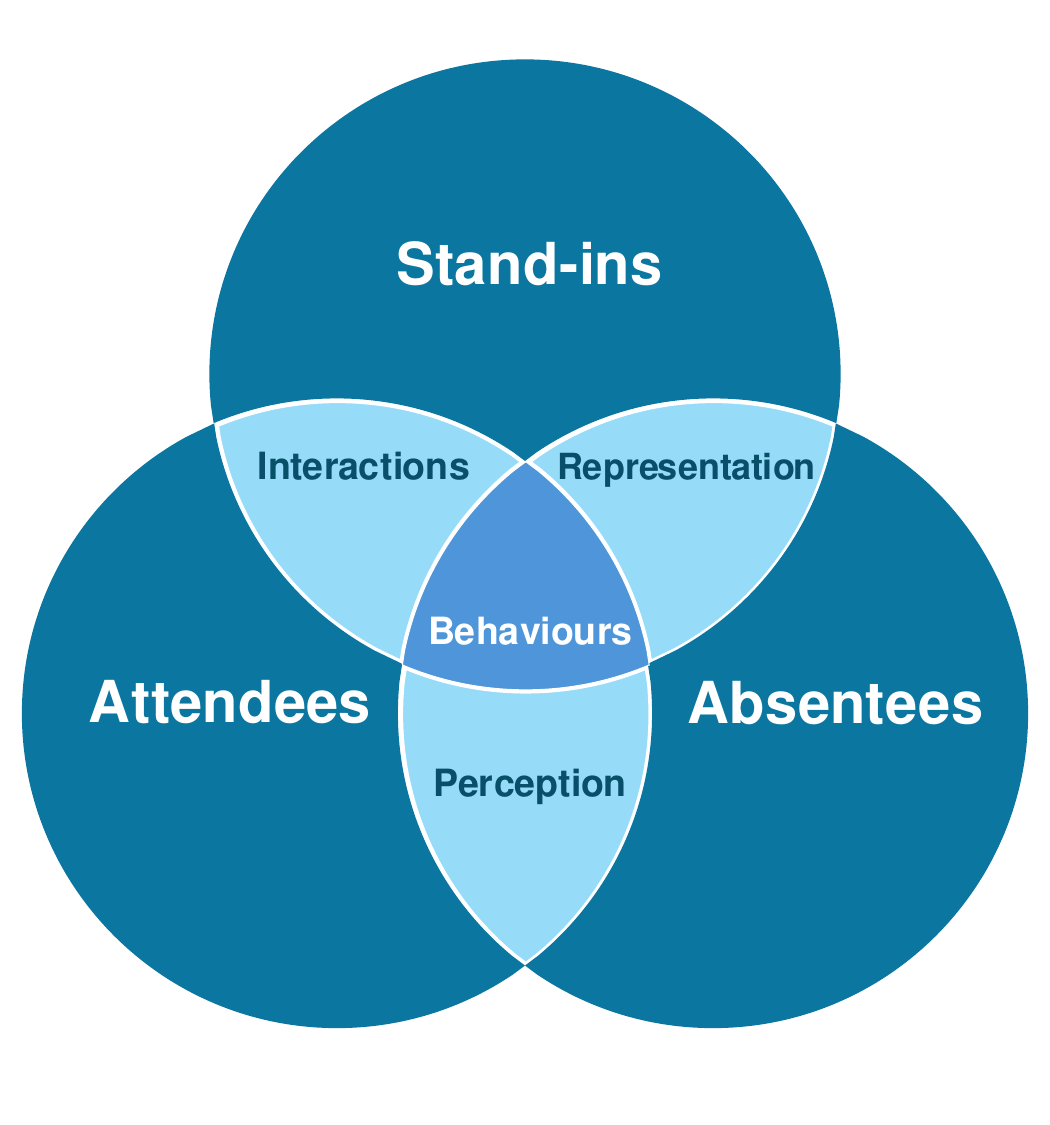}
    \caption{The dynamic interplay between Attendees, Absentees, and Stand-ins}
    \Description{This image depicts a Venn diagram illustrating the dynamic interplay among three entities: Attendees, Absentee, and Stand-in. The diagram consists of three overlapping circles, each labelled with one of these entities. The overlapping areas represent various elements of interaction and influence between these groups. At the intersection of all three circles, the term "Behaviours" is placed, indicating that behaviours are influenced by and impact all three categories. The overlap between Attendees and Stand-ins is labelled "Interactions," suggesting direct exchanges between these two entities. The overlap between Attendees and Absentees is labelled "Perception," pointing to how attendees perceive the absentee. Finally, the overlap between Absentee and Stand-in is marked "Representation," reflecting how the stand-in represents the absentee in interactions.}
    \label{fig:results-structure}
\end{figure}

We present our findings from Study 1 and 2 in two parts. First, we present results from a quantitative data analysis that show the differences in the perceived social presence between attendees and absentees from Iteration 1 and Iteration 2. Second, we present results from our observation and interview data set involving the three elements described in our findings: \textit{Attendees}, \textit{Absentees}, and \textit{Stand-ins} (\autoref{fig:results-structure}).
Our findings reveal the dynamic interplay between the three roles; the three \textit{interactions}, \textit{representations}, and \textit{perception} bridges connecting each role, while highlighting the shared \textit{behaviours} between all roles.

Initial interviews in both studies helped us understand participants' prior experiences of meetings where at least one participant was absent and the consequences of their absence. Participants conveyed two primary issues when this situation occurred:
(1) the absentee missed the opportunity to contribute when it was necessary, and 
(2) additional time was wasted discussing unresolved problems with the absentee after the meeting occurred.
When reflecting upon times when they were absent and how they caught up, the most common solutions were reading textual notes, talking to one of the attendees, or watching video recordings of the meeting. 

\subsection{Differences in the perception of the social presence (RQ1, RQ4) -- Quantitative Results} \label{section 7.1}
We used a Bayesian mixed ordinal regression model with a cumulative probit link built with \texttt{brms}~\cite{burkner2017brms} to analyse the questionnaire data. This method is widely used in HCI research, which offers the capacity to quantify uncertainty and the ability to facilitate future work to build upon it~\cite{li2025wice,li2025encumbrance}. These models estimate effects assuming a latent standard normal distribution, so the regression coefficients for binary variables can be interpreted as a measure of effect size similar to Cohen's d values. Full coefficient tables can be found in the Appendix. The Networked Minds questionnaire includes two parts: perception of self and perception of the other. We compared three levels of attendance: Attendee in Iteration 1 - real human (Study 1), Absentee in Iteration 1 - represented by the stand-in (Study 1), and Absentees in Iteration 2 - recorded avatar from Iteration 1 (Study 2). Our results show that, on average, in terms of the perception of self, attendees in Iteration 1 perceived the absentee in Iteration 1's social presence as -0.75 [-0.97, -0.61] standard units lower than the attendee in Iteration 1's social presence. Attendees in Iteration 2 perceived the absentees in Iteration 2's social presence as -0.16 [-0.48, 0.18] standard units lower than the attendee in Iteration 1's social presence. This finding is similar to the perception of the other, with a reported value of -0.73 [-0.86, -0.60] and -0.54 [-1.03, -0.04] (see Table \ref{tab:perception-social-presence}). 

Our self-designed questionnaires (Table~\ref{tab:self-designed-questions}) measured three constructs: \textit{Perceived contribution}, \textit{Inclusion}, and \textit{Appropriate behaviours} by comparing two states: before and after watching the recorded VR meeting. Our results show that participants felt the original absentee made more contributions before watching the recorded VR meeting with a difference of 0.64 [0.04, 1.24] (see Table \ref{tab:perception-contribution}).  Table \ref{tab:perception-inclusion} shows participants thought the original absentee felt less included before watching the recording than after watching the recording. As shown in Table \ref{tab:perception-behaviours}, participants rated their behaviour towards the original absentee as 0.49 [-0.11, 1.13] units more respectful before watching the recording than after. These results show how the activity of watching their own meetings affects participants' perceptions of their own behaviour.

\begin{figure}[h]
    \centering
    \begin{minipage}{0.48\textwidth}
        \centering
        \includegraphics[width=\textwidth]{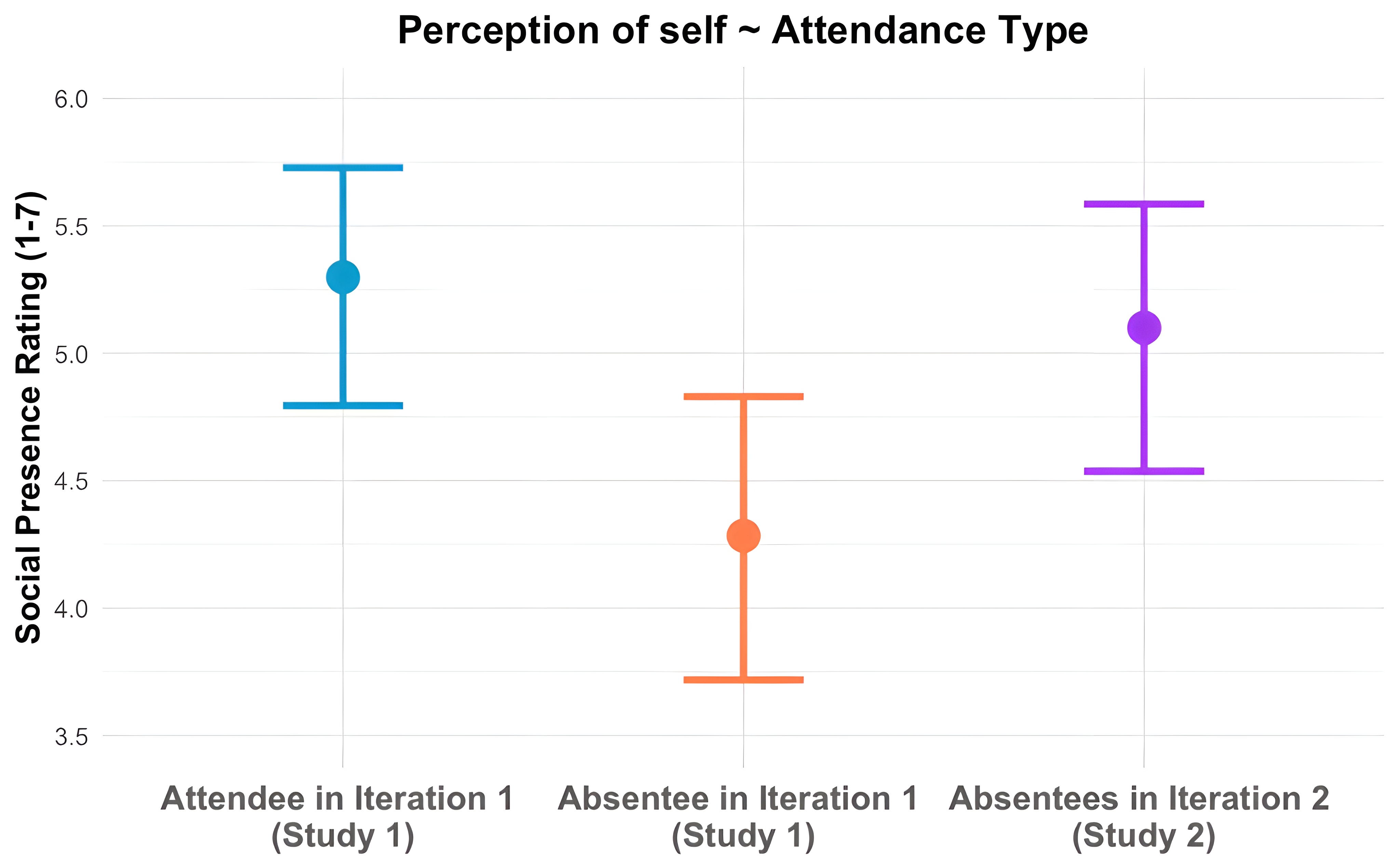}
    \end{minipage}
    \hfill
    \begin{minipage}{0.48\textwidth}
        \centering
        \includegraphics[width=\textwidth]{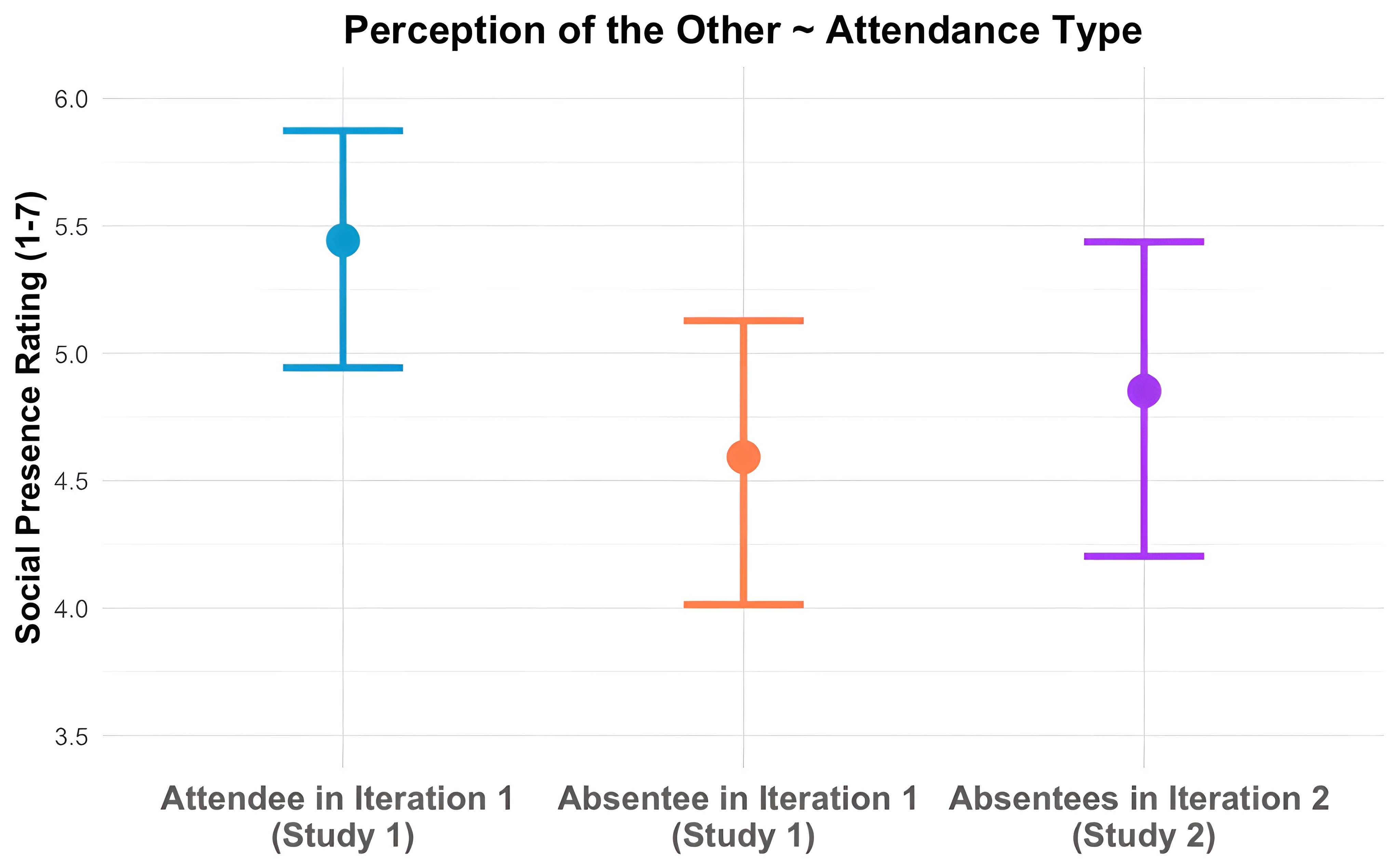}
    \end{minipage}
     \caption{Posterior estimates of mean social presence scores with 95$\%$ compatibility intervals. Scores relate to the perception of self (e.g. ``I often felt as if $<$other$>$ and I were in the same room together'') and of the other (e.g. ``I think $<$other$>$ often felt as if we were in the same room together''). The ``attendee in Iteration 1'' was the other participant in Study 1, the ``absentee in Iteration 1'' was the absentee represented by the stand-in, and the ``absentees in Iteration 2'' were the original attendees who recorded the meeting in Study 1. We found attendees in Iteration 1 (study 1) generally reported higher social presence scores compared to absentees in both iterations. We note that these are not 95\% confidence intervals, so no claims of statistical significance are to be derived from these.}
     \Description{This figure shows a posterior estimate of mean social presence scores with 95$\%$ compatibility intervals.}
    \label{fig:posterior-estimates}
\end{figure}

In summary, these findings reveal a medium to large gap in the ability of the stand-in to fully represent an absentee's social presence (see Figure~\ref{fig:posterior-estimates}). We speculate this gap is caused by the stand-in's limited responses and less human-like interaction performed during the study task, our qualitative analysis also supports this point. Additionally, even though the absentees in Iteration 2's behaviours and interactions were all derived from real participants in Iteration 1, they also cannot provide full social presence due to the lack of real-time interaction in recordings. These findings help us understand the basic gap between a stand-in and a real human. Future research may aim to reduce the gap by improving stand-in intelligence and interactions.

\begin{table*}[!htbp]
    \centering
    \caption{Summary of the cumulative probit model for perception of the attendee and absentee's social presence in Iteration 1 and Iteration 2: Social presence $\sim$ Attendance + (1|Participant) + (1|Questions) + (1|Meetings). We provide the posterior means of parameter estimates (Estimate), standard errors (Error), and the bounds of the 95\% compatibility interval (CI). All parameter estimates converged with an ESS well above 1000 and an R-hat of 1.00. Fixed effects are relative to the perceptions of the attendee in iteration 1. Other parameter estimates are shown in Table \ref{tab:perception-social-presence_full}.}
    \Description{This table presents a summary of a cumulative probit model analyzing the perception of social presence for attendees and absentees in a study. It displays parameter estimates, standard errors, and 95\% confidence intervals for two categories: "Perception of self" and "Perception of the other." The table is divided into three sections: Ordinal regression cut-points (τ[1] to τ[6]), Fixed Effects (Attendance), and Random Effects (Participant and Question). Each row shows the estimate, standard error, and confidence interval for the respective parameter.}
    \label{tab:perception-social-presence}
    \begin{tabularx}{\textwidth}{>{\raggedright\arraybackslash}p{3.3cm}XXXX}
    \toprule
    \multirow{2}{*}{\textbf{Parameter}} & \multicolumn{2}{c}{\textbf{Perception of self}} & \multicolumn{2}{c}{\textbf{Perception of the other}} \\
    \cmidrule(lr){2-3} \cmidrule(lr){4-5}
    & \textbf{Estimate (SD)} & \textbf{95\% CI} & \textbf{Estimate (SD)} & \textbf{95\% CI} \\ \midrule
    
    \multicolumn{5}{l}{\textbf{Fixed Effects}} \\
    Absentee in Iteration 1  & -0.75 (0.07) & [-0.97, -0.61] & -0.73 (0.07) & [-0.86, -0.60] \\ 
    Absentees in Iteration 2 & -0.16 (0.17) & [-0.48, 0.18] & -0.54 (0.25) & [-1.03, -0.04] \\  
    \bottomrule
    \end{tabularx}
\end{table*}

\subsection{Attendees' Perspective (Study 1)}
\subsubsection{Interactions with the stand-in (RQ1, RQ2)}
As the attendees in Iteration 1 interacted with the stand-in and became accustomed to it, they adapted their interaction strategies. 
Two changes in communication styles were observed: (1) from a social-oriented to a task-oriented communication style, and (2) from a task-oriented to a social-oriented communication style. In Embodied Conversational Agents (ECAs) studies, a social-oriented communication style is aimed to address emotional needs and enhance closeness that can be reflected in the use of personalized greetings or engaging in small talk \cite{advisor-communication-style, bass-managerial-applications, communication-style-eca}. On the other hand, a task-oriented communication style focuses on completing tasks efficiently.

When observing the transition from a social-oriented to a task-oriented communication style, initially, many participants treated the stand-in as a real human, \textit{``How are you doing?''(G10-B) ``Hi, how are you?''} while waving hands. G06-A also highlighted that \textit{``I looked at him (stand-in) and asked questions without saying `Hey Lee' ''} explaining that there was no need to say "Hey" to get their attention, as you already share a mutual gaze with the person. As the session continued, participants began to notice the stand-in's ability to answer simple questions, adapting their communication style. For example, G10-B shifted towards asking a more task-related question \textit{``What do you think about the food?''} after not receiving an adequate response to their greeting. 

For the transition from task-oriented to social-oriented communication style, after seeing the stand-in's verbal response with non-verbal cues, G11-A started asking more complicated and compounded questions, \textit{``Ohh, are you good at hiking? ... Do you have any recommendation?''} (G11-A). 

We posit the change was caused by the expectancy violation in which participants initially think the stand-in is a bot that can only provide binary responses~\cite{expectancy-violations}. However, once the stand-in begins to exhibit more lifelike behaviours, such as using deixis alongside its responses, the participant's initial expectation is violated, influencing their communication interaction. In the observed examples, this change caused participants to adjust their interaction strategies with the stand-in towards a more social-oriented communication style.

\subsubsection{Facilitating decision-making (RQ3)} \label{facilitating-decision-making}
Currently, when a team member misses a meeting, the only way to have their views shared is through the distribution of notes beforehand. This places a responsibility on those in the meeting to remember to raise and discuss the views of the absentee. In contrast, the design of the stand-in reduces the onus on attendees to remember to include these views, allowing for key information to be interjected into the conversation in real time. This design results in the views of the absentee having a positive impact on decision-making during the meeting, a sentiment supported by G05-A:

\begin{quote}
    \textit{``In the typical way, when someone cannot attend, we just ignore them. Maybe we won't consider their thoughts. But in this (meeting with stand-in), we can ask them and get information immediately.''}
\end{quote}

As the stand-in could provide conversational responses based on verbal prompts, the absentee's views were incorporated into the meeting and the decision-making process. The ability of the agent to provide timely information with an understanding of the local context created a more efficient and accessible information retrieval process, a finding also highlighted in Rhodes and Maes's \textit{JITR} agent~\cite{jit-agent}.

Additionally, the stand-in's body language and clear responses (e.g., pointing to objects, gesturing alongside utterances) reinforced the absentee's preferences, which affected the final decision made by the attendees, \textit{``Lee's (stand-in) opinion influenced our decision''}(G06-B), \textit{``When I saw his body language [pointing at beef noodles] and saw beef noodles ... I changed my first choice from pizza to beef noodles''} (G14-A). From the Iteration 1 task results, 14 out of 15 groups took the absentee's preferences into account in their final decision. This finding demonstrates how the presence of a stand-in can be helpful in contributing towards the decision-making process during meetings.

\subsubsection{Absentee's social presence (RQ1)}
The stand-in's responses with non-verbal and verbal cues increased the attendees' perception of the stand-in as an actual participant.
73\% of participants reported that they perceived three people during the meeting; this includes two live \textit{attending} participants plus the stand-in. 
Just the presence of the stand-in increased this perception, \textit{``because I saw him (stand-in), he was just there and moving''} (G15-A); \textit{``it's like a visual thing that it (stand-in) reminds you there are three people in this room''} (G09-B). Furthermore, this perception only increased when the stand-in exhibited active listening and provided responses as part of the ongoing conversation, \textit{``...because it (stand-in) can answer some questions''} (G01-A); \textit{``(There were) three (users) because I will involve the stand-in's response in my decision making''} (G02-B); \textit{``three ... all the time, because it can speak''} (G12-A). A reduction in negative feelings and complaints towards an absentee was also reported by participants, highlighting that compared to Zoom, at least the absentee's stand-in was present and contributed towards the meeting discussion~\cite{equal-participation}. The stand-in's use of body language alongside their verbal responses helped increase the perception that they were in a real face-to-face meeting with the absentee.

Eleven participants reported that the number of people they perceived during the meeting changed from two to three because of the stand-in's responses in the conversation as G07-A highlighted \textit{``if the stand-in gives the response, (it) sometimes (feels like) three people''}; (G13-A) \textit{``before Lee (stand-in) could respond, it was two (people), however after Lee responded, it became three''}.

However, four participants reported they perceived the stand-in as neither fully absent nor fully present. The presence of the stand-in and its ability to respond to participants' questions made it difficult to ignore. Nevertheless, the stand-in's inability to answer more sophisticated questions and repeat answers made participants feel it was still not an actual human, \textit{``because he (stand-in) can only answer questions instead of discussing with us''} (G05-B), and \textit{``A stand-in is not good enough to make me think he is a real person''} (G14-A).  This result shows that although the stand-in's responses helped increase the attendees' perception of the absentee's social presence, they cannot fully represent the absentee. This also aligns with our quantitative results.

\subsection{Attendees’ Self-Reflection from the Absentee’s Perspective  (Study 1)}

\subsubsection{Behaviour change}
After watching the recorded meeting, participants reflected on their desire to improve future interactions with a stand-in.
In our study design, Iteration 2 tasked participants with watching their recorded meeting from Iteration 1. This gave each participant a unique opportunity to reflect on how their own behaviours in the meeting would be perceived by an absentee. After reviewing their meeting, 
87\% of participants expressed their desire to change their behaviour towards the stand-in, \textit{``ask questions [more] loudly''} (G11-A); \textit{``interact with Lee [the Stand-in] more''} (G13-B); \textit{``ask more questions and pay close[r] attention to Lee''}(G14-B); \textit{``add more gestures''}(G09-B). We speculate the desire to improve their future interactions is caused by the discrepancy between participants' internal intentions and external interpretations. For example, in Iteration 1, participants noted their motivation to seek the absentee's opinion, feeling that they were adequately involving the absentee in the conversation. However, during the experience in Iteration~2, participants realised that their behaviour (as an attendee in Iteration 1) commonly ignored the absentee and visibly showed dissatisfaction when not receiving a suitable response, with these behaviours creating a negative experience for the absentee. This also shows how the effectiveness of asynchronous meetings depends on long-term experience with such a workflow, as collaborators can benefit from experiencing both sides of these interactions.

\subsection{Absentees' Perspective (Study 2)}
During Study 2, each participant watched two different meetings recorded in Study 1 from an absentee's perspective. Our findings in Study 2 validate Study 1 Task 2's results that participants reported positive feelings towards being included when watching the recording. Additionally, Study 2's qualitative results also help us understand what factors made participants feel included and not included. 
\subsubsection{Factors made absentees feel included (RQ4, RQ5)}
We identified two factors (\textit{Social interactions} and \textit{Attention}) that made participants feel included when watching the recording from the absentee's perspective. \textit{Social interactions}, 66$\%$ of participants expressed that when absentees in Iteration 2 tried to ask them questions made them feel they were part of the conversation, \textit{``One thing was when they (absentees in Iteration 2) asked me questions directly, then I felt included''} (P01); \textit{``They (absentees in Iteration 2) just directly ask me''} (P11); \textit{``They (absentees in Iteration 2) have a really good turn-taking strategy''} (P05); \textit{``I think both of them asked me quite a few times. Like, what do you think...''} (P07). These social interactions resulted in attendees in Iteration 2 feeling that their opinions were being taken into account, even though they were not physically present in the meeting. 

Further, similar to the results from Study 1, participants in Study 2 also reported their experience in Iteration 2 to be similar to that of attending a live meeting, even though they were consciously aware that it was a recording of a past meeting. Participants expressed that the feature of allowing them to record comments and respond when watching the recording made them feel included as participating in a real meeting, \textit{``because I can leave some comments then I feel kind of included.''} (P06); \textit{``Like we are in the same space, like a live meeting''} (P03); 
\textit{``I think I didn't feel like absent like it felt it was happening in front of me.''} (P15). In the recording, the participant experienced the meeting through the perspective of the stand-in. This allowed them to have more social interactions, such as maintaining eye contact throughout the conversation and waiting for an opportunity to interject with their own responses. From our analysis, it was noted that 73$\%$ of participants left comments while watching the recording as if they were participating in the live meeting.

\textit{Attention}, non-verbal cues such as facing towards participants, eye contact, and listening behaviours made participants feel absentees in Iteration 2 were paying attention to them. As participants reported: \textit{``The non-verbal communication towards me. And looking at me''} (P02); \textit{``They (absentees in Iteration 2) were paying attention to me, they wanted to interact and talk to me''} (P03); \textit{``They (absentees in Iteration 2) actually listened to my opinion and included me into their decision making''} (P09). The attention focused on the stand-in made participants feel included as genuine members of the meeting and is likely a direct consequence of the embodied nature of the stand-in. 

\subsubsection{Factors made absentees feel not included (RQ4, RQ5)}
During Study 2, we also noticed participants did not always feel included when watching the recording, which aligns with our findings in Study 1. We identified two factors (\textit{Discrepancy in responses} and \textit{Lack of natural interaction}) that made participants feel excluded.  
\textit{Discrepancy in responses}, given the meeting had already taken place, several participants raised the issue of feeling their current interactions have no impact on changing the meeting outcomes and that their responses end up feeling unhelpful, \textit{``I expressed my preferences, they (absentees in Iteration 2) sort of immediately said something different''} (P07); \textit{``I said 'I'm not good at swimming' and they still said go to Manly beach''} (P09); \textit{``I left a comment but they just kept talking and this made feel not that included''} (P06). Due to the current immutable nature of the recording, the attendee's responses in Iteration 2 were not registered in the ongoing conversation. This limitation resulted in making participants feel weird and excluded, especially when the responses between the stand-in and the attendee differed, with the conversation always continuing on the basis of the stand-in's response.

\textit{Lack of natural interaction}, several participants also reported the way how absentees in Iteration 2 interacted with them was not natural as human-to-human interaction. Participants expressed they were treated more like a voice assistant because whenever absentees in Iteration 2 tried to ask questions to participants, they always started with words \textit{``Hey Lee...''}, which made them feel like trigger words for voice assistant. As P02 highlighted \textit{``I felt a bit like being a voice assistant. In a normal meeting, you won't say like this (Hey Lee)''}. In addition, participants also reported sometimes absentees in Iteration 2 tended to discuss by themselves and ask participants' opinions until the end of the meeting, which made them feel like a bystander and the meeting conversation was not natural, \textit{``I didn't feel like it felt natural mostly when in the end, they asked me 'What do you think'''} (P01).

\subsubsection{Catching up on missed meetings (RQ4)}
Participants expressed they would prefer to watch the embodied recording than read notes or watch a Zoom recording if they needed to understand the reasoning behind how and why the decision was decided upon, including understanding the emotions and attitudes of attendees during this process~\cite{social-roles-emotions}, as participants reported \textit{``If I can watch the VR recording and then I can understand what was exactly going on when they had this conversation. I want to know their interaction''} (P05); \textit{``I guess I can catch up more details of this meeting, like how they made their decisions and I can know what they were worried about''} (P11). However, for non-important meetings or during busy times, participants would still prefer to get a summary of the meeting for better efficiency.

In Study 2, participants also highlighted the key benefits of VR recordings over traditional videoconferencing approaches, which were non-verbal cues and a stronger sense of contributing to the discussion. VR recordings provided a more immersive environment with embodied avatars that allowed participants to perceive richer non-verbal behaviours than Zoom recordings, as participants highlighted \textit{``I would say turning towards me is definitely a cue that you don't have recorded in Zoom''} (P02); \textit{``(In VR recordings) I have received information from both verbal (cues) and non-verbal (cues)''} (P03); \textit{``There are more interactions (in VR recordings) between people''} (P09). In our prototype, participants were allowed to record responses and comments while watching the recording. This feature helped participants develop a stronger sense of participating in the meeting, \textit{``The biggest difference (compared to Zoom) is I'm kind of contributing to the meeting''} (P04); \textit{``This (VR recording) made me feel engaged because of the view and I can leave my comments directly to their response.''} (P06); \textit{``I'm just taking their conclusions, and I cannot express my opinions. It (Zoom recording) is just one way (communication).''} (P09). However, it was noted that the embodied nature of the communication helped participants stay engaged, the inability to quickly seek to key parts of the meeting or review a summary of the meeting were commonly requested system features to assist in overall time management.

\subsection{Expectations of the Stand-in (Studies 1 and 2)}
In both studies, participants were asked to describe their expectations of the stand-in. In this section, we present findings related to how participants wanted their stand-in to represent themselves (representation) and how the stand-in should be engaged in the meeting on the absentee's behalf (engagement).
\subsubsection{Representation}
Across both studies, 17 participants expressed the preference for a more realistic avatar (e.g. Avaturn\footnote{Avaturn: \url{https://avaturn.me/}}) while 25 for a non-realistic avatar (e.g. ReadyPlayerMe\footnote{ReadyPlayerMe: \url{https://readyplayer.me/}} or Meta Avatars\footnote{Meta Avatars: \url{https://www.meta.com/au/avatars/}}) to be their stand-in. Participants who preferred realistic avatars explained that they felt it could make other members have a stronger sense of being in a real meeting as G15-B highlighted \textit{``It looks like more realistic and I feel like I'm in the real meeting''}; \textit{``Because it (Avaturn) resembles more of me. It gives my friends more confidence and makes them willing to talk with me.''} (P04). The realistic avatar is also helpful for attendees to recognise active speakers faster, as G14-B reported \textit{``For the left one (Avaturn), I can quickly recognise who is speaking''} and P03 highlighted \textit{``That's why I like Avaturn, people can recognise me''}. Participants who preferred non-realistic avatars were mainly concerned about how life-like avatars would make them feel the stand-in could appropriate more authority as G08-B reported 
\begin{quote}
    \textit{``I feel like the Avaturn (avatar) is like pretending to be you, but it's not actually being you, which I find a bit creepy''}
\end{quote}
Additionally, several participants would choose different avatars depending on the type of meeting the stand-in would attend. Professional meetings called for realistic avatars, whereas social meetings called for non-realistic or customisable ones. The stand-in's voice was also highlighted as an important factor that needs to be considered. While there was more freedom to change an avatar's visual appearance, the stand-in should still sound like the absentee it is representing, \textit{``I will talk to the AI and it will learn from my voice, my actions...''}(G06-A); \textit{``I probably want my stand-in to sound or speak more like me''} (P14).

\subsubsection{Engagement}
Active participation plays a crucial role in the meeting; it requires participants to provide input or share thoughts proactively without being addressed. Many participants reported that the stand-in only participated in the conversation when it was addressed. Sometimes the stand-in was viewed more as an observer than a participant, \textit{``I think there were three (people), but at the beginning of the meeting, I just think he (stand-in) was like a listener''} (G06-B). In both studies' interviews, several participants expressed their expectation that their stand-in should be more proactive in the discussion without being triggered by questions. 

The representation of the stand-in was not powered by AI and could not answer complicated questions. This limitation resulted in participants expecting their stand-in to have more intelligent responses, \textit{``It can behave smarter''} (G12-B). Additionally, the inadequate responses provided by the stand-in also made absentees have to adjust their response when watching the recording as P5 highlighted \textit{``Mainly when I realize the stand-in didn't do well to some questions, that's where I had to add comments''}. Furthermore, the lack of more advanced communication skills created a desire for stand-in's to defer decision making as P10 reported \textit{``If the stand-in is not 100$\%$ sure what I would respond, I hope the stand-in just say 'I will reply later' like this'' and } G08-A highlighted:
\begin{quote}
\textit{``If it's a really important decision making, I hope my stand-in tell me first and don't make any decision first, just because I don't know what it will say.''} 
\end{quote}

Twenty-five participants reported they wanted their stand-in to learn the behaviours from their past meetings (e.g. Zoom recordings). G02-A also mentioned that it would be good to provide feedback to the stand-in and train the stand-in to behave more appropriately. Customized avatars are widely used in video games to allow players to customise their appearance. In this study, several participants reported they prefer to configure stand-in behaviours. In a similar fashion to video games, they could manually customize specific fields for stand-in behaviours. For example, allowing users to adjust stand-in's behaviours to be more casual or professional using a slider depending on what meeting the stand-in will attend. Additionally, participants also expressed the expectation of having different stand-in profiles to fit different meeting settings and purposes, as P05 reported 
\begin{quote}
    \textit{``If we can set different profiles for my stand-in. For example, I will set up a stand-in as my academic profile and I will set a stand-in (profile) for my friends.''}
\end{quote}

\section{Discussion}
In this section, we discuss the future research focus on expanding on the concept of SEAM from two aspects: (1) the \textit{design considerations} related to system improvement and key functionality support; (2) \textit{trust, privacy and ethics}---three key factors that participants raised during the interviews; and (3) \textit{LLM Capabilities and limitations} discussing both the potential of LLMs to enhance contextual understanding and the current shortcomings such as limited memory and inconsistent behaviour over time.

\subsection{Design considerations}
After trying the prototype from both the attendee's and absentee's perspectives, participants provided many insights into improvements to the SEAM experience. We note two common themes: stand-in design and playback features. The stand-in's design directly influences the attendees' experience in the meeting, while the playback feature is an essential part of allowing the absentees to catch up on missed meetings. 

\subsubsection{Stand-in}
From our results, participants seemed to care about their stand-in appearance and behaviours. In addition, participants also mentioned they wanted their stand-in to make a good impression as the stand-in's behaviours represent their own behaviours. However, a stand-in should express a disclaimer before the meeting starts, making it transparent that its contributions may or may not reflect the absentee's thoughts. This could reduce the absentee's concerns about how the stand-in would behave and what it might say during the meeting.

Future stand-in designs should consider adding more flexibility to the stand-in's appearance and behaviour design. It should give users more ways to configure their stand-in and offer more options to fit in different use cases. For appearance design, as results show, not all participants prefer having a realistic representation as their stand-in. Too realistic avatars could cause an uncanny valley effect \cite{uncanny-valley}, leading to an eerie sensation that could affect the overall meeting experience. In contrast, with non-realistic representation, participants would have less pressure on their self-image. It is also unclear whether using a realistic or non-realistic stand-in could cause the Proteus effect \cite{proteus-effect}, a phenomenon where the characteristics of a virtual avatar influence the user’s behaviour and self-perception. In this case, absentees using non-realistic avatars for their stand-in might feel more comfortable and less concerned about how others perceive them because the stand-in does not closely resemble them.

In terms of behaviour design, future systems could add more options for users to adjust the level of engagement of stand-ins. For example, if the absentee's role in a meeting is as an equal contributor, their stand-in would behave more actively and provide insights more frequently; if the absentee's role is as an audience member, their stand-in would behave more passively. Future work should investigate the levels of engagement that could be configured for such behaviour. Moreover, this study only evaluated SEAM within a small-group setting, featuring only a single stand-in. If the meeting group size expands such that numerous participants dispatch their stand-ins, how should these stand-ins act? Will this lead to enhanced efficiency or result in disorder? Both agent-to-agent and agent-to-human interactions become increasingly more complex, leading to interesting questions for future work.

Participants reported how the stand-in's responses influenced their decisions in the meeting (see section \ref{facilitating-decision-making}). However, one major complaint from participants was that the stand-in could not answer complex questions. Additionally, our results indicate that participants treated other attendees and the stand-in differently. Participants treated the stand-in more patiently (e.g. waiting a longer time for a response) as they understood the stand-in might need some time to process their questions and respond as they would with the usual voice assistant. Thus, it is important to leverage AI and LLMs to make stand-ins more intelligent and give them the ability to negotiate or interject appropriately during the meeting. To provide a better response on behalf of the absentee, a stand-in may need to involve the absentee's personality and personal information. Providing reasonable and almost real-time responses requires the stand-in to have a good contextual understanding.

In addition, based on our own first-hand experience of using the prototype for two months (April and May 2024) over eleven asynchronous meetings, we found that the absentee's embodied stand-in helped attendees perceive the absentee's social presence and led attendees to seek the absentee's opinions during the meeting. However, we still used Zoom when digital content sharing was necessary or a VR/MR headset was not available. Thus, future studies may explore whether and how the concept of SEAM can be integrated into multi-device workflows, such as combining MR and videoconferencing.

\subsubsection{Playback}
Previous studies have explored different ways to improve an absentee's experience of catching up on missed meetings. For example, Wang et al. have recognised how spatial accommodation and gaze affect perceived attention for new users who watch recorded meetings \cite{socially-late}. Girgensohn et al. explored a hyperlink-based asynchronous meeting concept that allows users to navigate meeting topics in a linked chain \cite{hypermeeting-async-meetings}. 

However, as in a SEAM, the absentee and their stand-in could be considered as two entities. As such, it is unclear the best way for absentees to watch recorded meetings. Should absentees watch from a third- or first-person perspective? If it is from a first-person perspective, how should the system illustrate stand-in responses? If it is from a third-person perspective (e.g. watch as a fourth character), how should future Iterations perceive the behaviours generated in this Iteration? As the number of Iterations grows, there will be more and more versions of characters leading to scalability issues. Those questions must be considered in future SEAM systems, as the meeting could continue if an absentee opens a new conversation while watching the recording. Additionally, controls are required to allow users to easily navigate and change recording speed to quickly get key results from the meeting. A potential solution could be post-processing the recorded meeting and creating event anchors. While absentees watch the recording, key events with concise context summary for the event could be displayed, so users can easily navigate to events that they are interested in.
With post-processing, the stand-in could also notify the absentee about the summary of the meeting and what the stand-in responded to during the meeting, allowing the absentee to revise any incorrect response that the stand-in made. Additionally, future systems also need to consider how absentees leave comments in the recordings. In our prototype, participants had to press a button on their controller to pause the ongoing recording and record their responses. However, when pausing the recordings, every recorded avatar freezes, which makes the experience unnatural. Thus, the future system should also consider enabling stand-in behaviours for absentees in Iteration 2, such that during pauses in the recording, the stand-in can enact listening behaviours.

\subsection{Trust, Privacy and Ethics}
\subsubsection{Trust}
Our SEAM vision raises complex questions about trust. Can absentees trust their stand-in's responses in the meeting? Can attendees trust the responses of stand-ins? Though the absentee provides some inputs for their stand-in, this delegation of attendance could lead to perceptions of a loss of control and issues around accountability and transparency. For example, in reporting on SEAMs, are stand-in behaviours and responses attributed to the absentee or the attendee? Are errors in generated gestures or responses attributed to the system powering the stand-in, or do they reflect on the absentee?

While these questions undoubtedly require further work, providing transparent control systems is a crucial first step towards allowing absentees to trust their stand-in or allowing attendees to control an overzealous stand-in.

One participant mentioned that they wanted to double-check with the absentee after the meeting, even though the stand-in provided key information that helped their decision-making. This participant felt the stand-in behaved unintelligently and did not fully trust what the stand-in said. 

Future work should consider how to design a mechanism to make stand-ins more trustworthy. As an example, the system could display symbols to indicate which stand-in responses are verified by the absentee and which are generated by the stand-in itself.

\subsubsection{Privacy}
As AI and LLMs have developed rapidly in recent years, privacy is one of the biggest concerns that users are worried about. Many AI practical applications also support personalised chat-bots or agents to meet users' needs. However, there is a trade-off between privacy and AI models --- how much personal information would you be willing to provide to make an AI more personalised? This trade-off can also apply to stand-in design. One interesting observation from our study results is that participants showed very different attitudes regarding sharing their personal information to train their own stand-in. One participant mentioned he would be willing to allow the stand-in to have full access to his personal information so that the stand-in can learn more from them and act more like them with personality. In contrast, some participants strongly expressed that they would not share any personal information with the stand-in because they worried that other attendees could launch "prompt injection" to encourage the stand-in to leak their sensitive information.

It is important for future systems to integrate not only technical protections but also ethical models to address the risks of misrepresentation and unintended data exposure. When an AI stand-in speaks or acts on behalf of the absentee, there is a risk that it expresses things not endorsed by the absentee, due to reasons such as imperfect training data, biases in language models, or prompts from other participants. Such issues can lead to reputational harm, a loss of trust, or social misunderstanding. Therefore, systems should adopt transparent mechanisms for disclosure. For example, it could include tagging AI-generated content, maintaining conversation logs for user review, or setting pre-established boundaries for how the stand-in should behave. Consideration should also be given to consent processes and interfaces that simplify the inspection and correction of the stand-in’s knowledge or expressions by the absentees. Implementing mechanisms for real-time intervention, or ‘safe words’ to pause actions of the stand-in can further guard against misuse. These protective measures are essential to ensure that AI-mediated stand-in remains trustworthy, reliable, and accountable.

\subsubsection{Ethics}
Although stand-ins could facilitate meetings and make absentees feel more included, they also bring ethical challenges, such as autonomy and long-term implications. 

For autonomy, the extent to which stand-ins should respect absentee autonomy is unclear. If everything produced by the stand-in needs to be controlled, the absentee will have to provide so much information that the perceived value of the stand-in may be significantly reduced. 

However, without any control from the absentee, stand-ins may cause unexpected consequences. Where prior work has explored agent design, agents are independent entities, whereas stand-ins are representations of an absentee. As such, it is unclear whether prior knowledge in agent design necessarily applies to stand-ins. 

For long-term implications, it is also unclear if an absentee's stand-in attends the meeting over multiple sessions, how other attendees would perceive the absentee, and if anything would change in the attendee's mental model. This scenario could happen if collaborators work in different time zones and cannot schedule a time that suits everyone. Those questions open new opportunities for researching SEAMs.

\subsection{LLM Capabilities and Limitations}
Large Language Models (LLMs) enable a stand-in to generate reasonable responses in meetings and mimic the absentee's communication style. However, they encounter significant challenges. LLMs often lack access to full contextual understanding such as the absentee's intent, history, or organisational insights, which can lead to inaccurate expressions. It is also challenging for LLMs to capture the personality traits, tone, and subtle communication styles, making it hard to genuinely represent the absentee. Additionally, current LLMs are not well-integrated with real-time generative animation, hindering the expressiveness and embodiment required for immersive settings like VR. Lacking visual understanding and continuity of memory, these stand-in can behave inconsistent over time. Future work should investigate integrating LLMs with structured memory and multimodal interactions to develop more intelligent and reliable stand-ins.

\section{Personalised LLM Stand-in System}\label{new-system}
In this section, we present the personalised LLM stand-in prototype system, extending SEAM from a earlier probe into a full functional system. Compared to the WoZ prototype used in Study 1, the personalised LLM stand-in system represents an evolution from a non-AI stand-in to a fully functional AI-powered stand-in. Whereas the original prototype required manual triggering of pre-recorded responses and exhibited only basic listening behaviours, the enhanced system enables autonomous stand-in interactions powered by live LLM reasoning, speech-to-text, and text-to-speech pipelines. It also supports persistent user profiles and a web dashboard for configuring stand-ins and agendas. Together, these enhancements move SEAM from a concept exploration towards a real-time system that supports personalised, embodied asynchronous meetings without human intervention.
\subsection{System architecture}
\begin{figure*}
    \centering
    \includegraphics[width=1\textwidth]{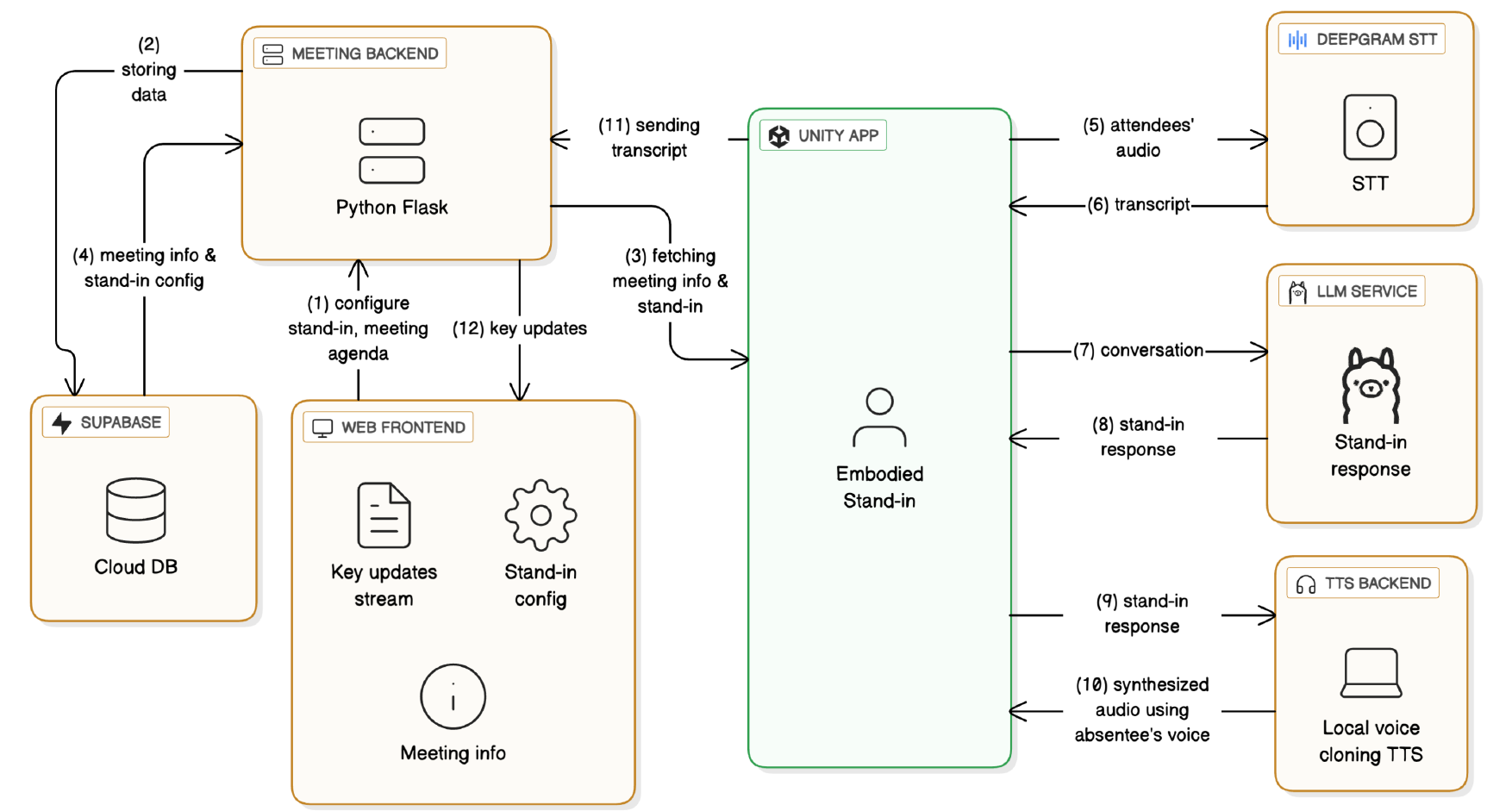}
    \caption{System architecture diagram: (1) The absentee first needs to provide their personal information (e.g. background, personality) in the system, and their stand-in will use this info. Then, the absentee needs to configure their stand-in regarding notes on each agenda item. (2) Once the backend receives the absentee's provided information, it will save all data into the database. (3) The stand-in will start the VR meeting, and during the initialisation stage, it will fetch meeting info and stand-in configuration (e.g. personal information, agenda items) from the backend. (4) The backend will fetch the requested data from the database. (5,6) When other attendees join the meeting and start the conversation, all audio will be sent to Deepgram, a speech-to-text (STT) service provider, to get the transcript. (7,8) The transcript will then be sent to the LLM service (using either OpenAI api or local Ollama api) to get the stand-in's response. (9,10) Once the stand-in's response is received, it will be sent to the on-device text-to-speech (TTS) backend. The backend will use the absentee's provided voice to synthesise audio and send it back to the stand-in in Unity. Then the stand-in will play this audio. (11,12) For all transcripts, we also send them to the meeting backend for the future partial participation feature.}
    \label{fig:seam-arch}
\end{figure*}

The full prototype system now consists of seven primary components, as illustrated in Figure~\ref{fig:seam-arch}.

\begin{itemize}
    \item \textit{Unity Application~\footnote{\url{https://unity.com}}}: This is the central component for the embodied stand-in, and the system was built using Unity 6.
    \item \textit{Meeting Backend~\footnote{\url{https://flask.palletsprojects.com}}}: A Python Flask backend manages meeting-related functions.
    \item \textit{Database~\footnote{\url{https://supabase.com}}}: Supabase using PostgreSQL serves as both the database and authentication systems.
    \item \textit{Web Application~\footnote{\url{https://react.dev}}}: A SEAM dashboard React web app allows users to schedule meetings and configure their personal profiles and stand-ins.
    \item \textit{Speech-to-text~\footnote{\url{https://deepgram.com}}}: Deepgram facilitates speech-to-text services.
    \item \textit{LLM Service~\footnote{\url{https://ollama.com}, \url{https://platform.openai.com}}}: Gemma3 in Ollama exposed through an API endpoint was used as a local LLM. Remote inference is also supported with OpenAI's GPT API.
    \item \textit{Text-to-speech~\footnote{\url{https://pinokio.co}}}: Code from Pinokio was adapted with the e5-tts library and established a Flask server to process text inputs and return synthesised audio, which the embodied stand-in then plays.
\end{itemize}

\subsection{LLM-powered Stand-in}
Before sending a stand-in to the meeting, each user involved in a SEAM submits a 10-second voice sample in WAV format for voice synthesis and creates an account on the SEAM dashboard. Additionally, users can configure their stand-in's background and personality, and arrange meetings by specifying the agenda items. Users are represented by avatars created in Ready Player Me studio, which generates 3D characters from a single submitted profile photo.

\begin{figure*}[ht]
    \centering
    \begin{subfigure}[t]{0.32\textwidth}
        \centering
        \includegraphics[width=\textwidth]{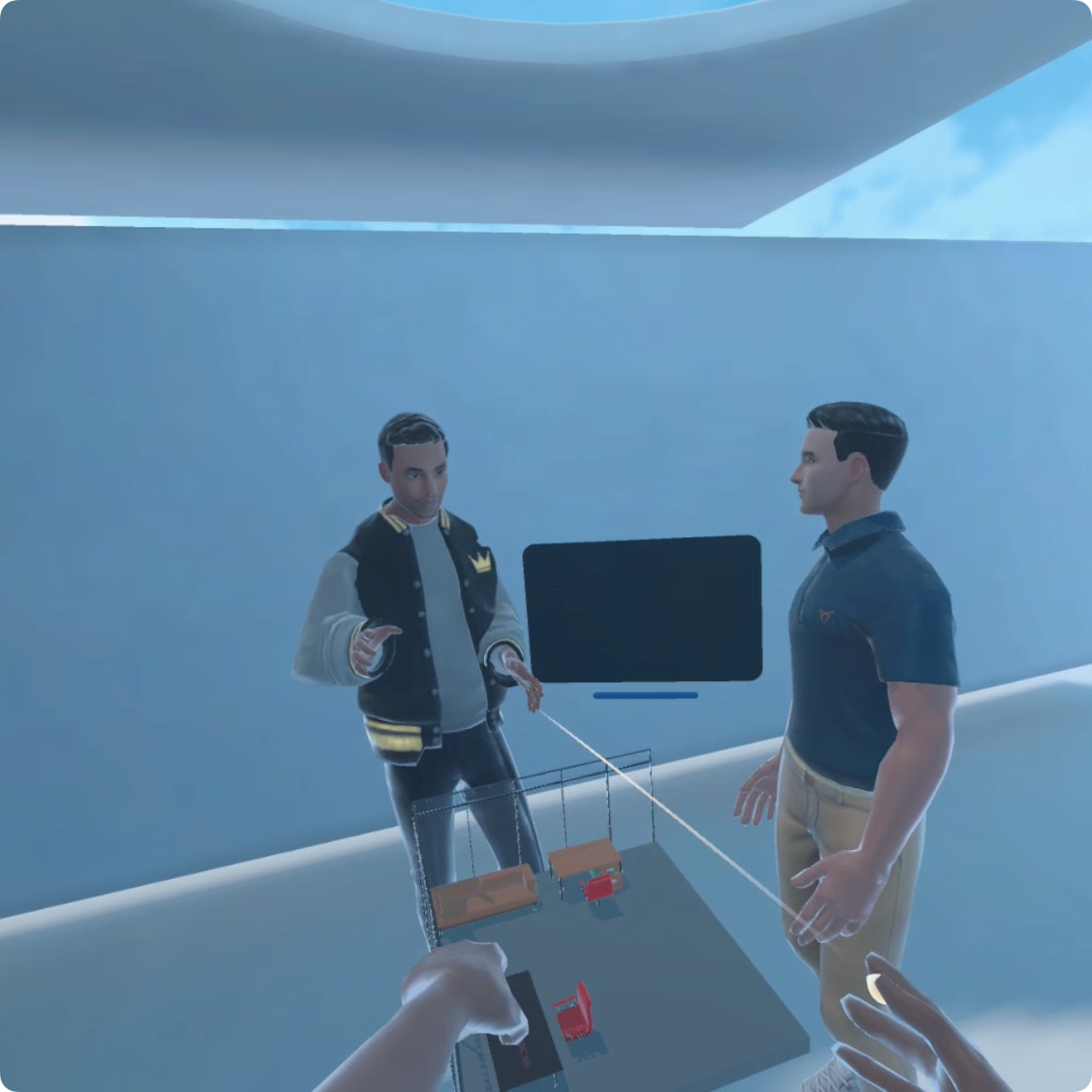}
        \caption{Stand-in listening}
        \label{fig:stand-in-1}
    \end{subfigure}
    \hfill
    \begin{subfigure}[t]{0.32\textwidth}
        \centering
        \includegraphics[width=\textwidth]{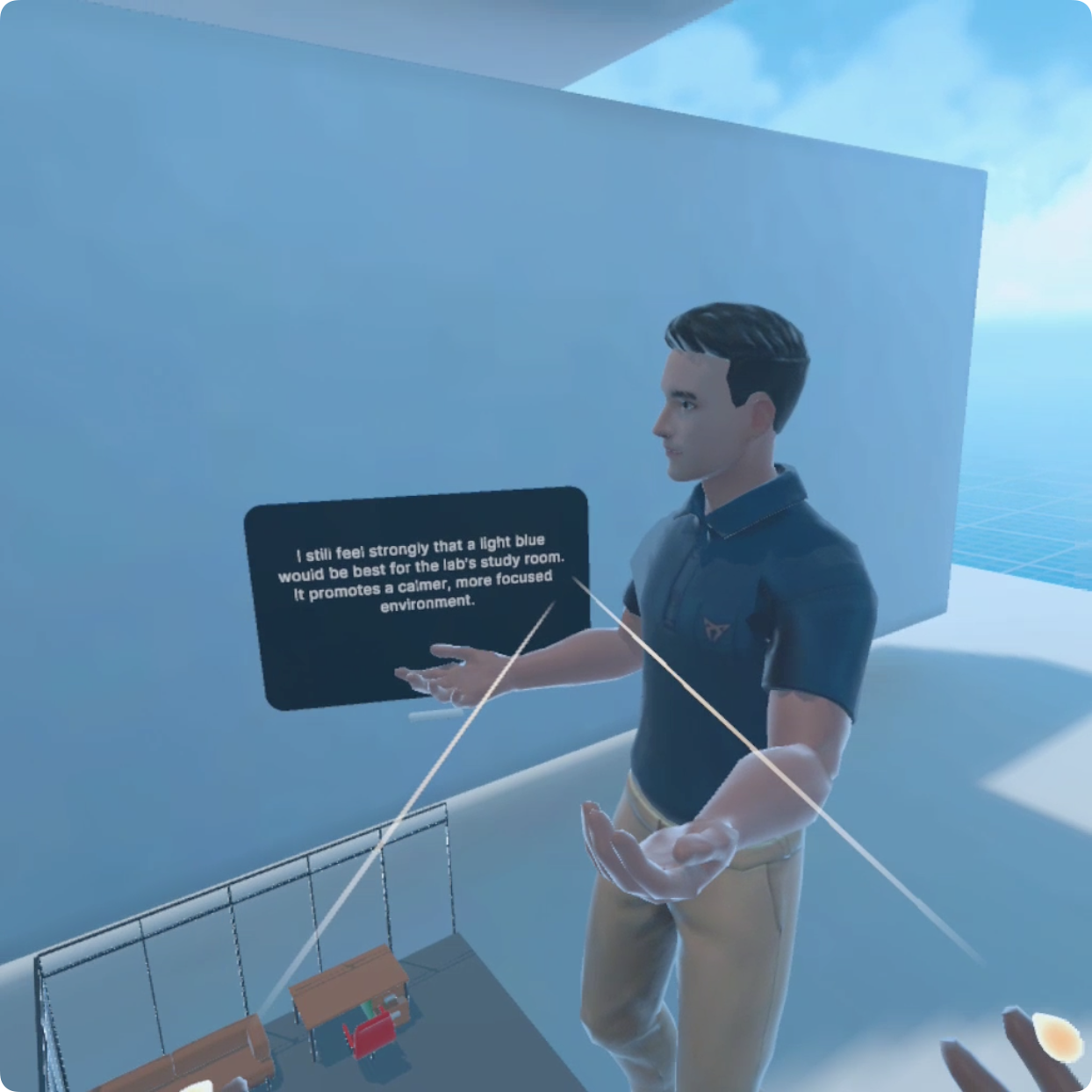}
        \caption{Stand-in actively participates in the conversation}
        \label{fig:stand-in-2}
    \end{subfigure}
    \hfill
    \begin{subfigure}[t]{0.32\textwidth}
        \centering
        \includegraphics[width=\textwidth]{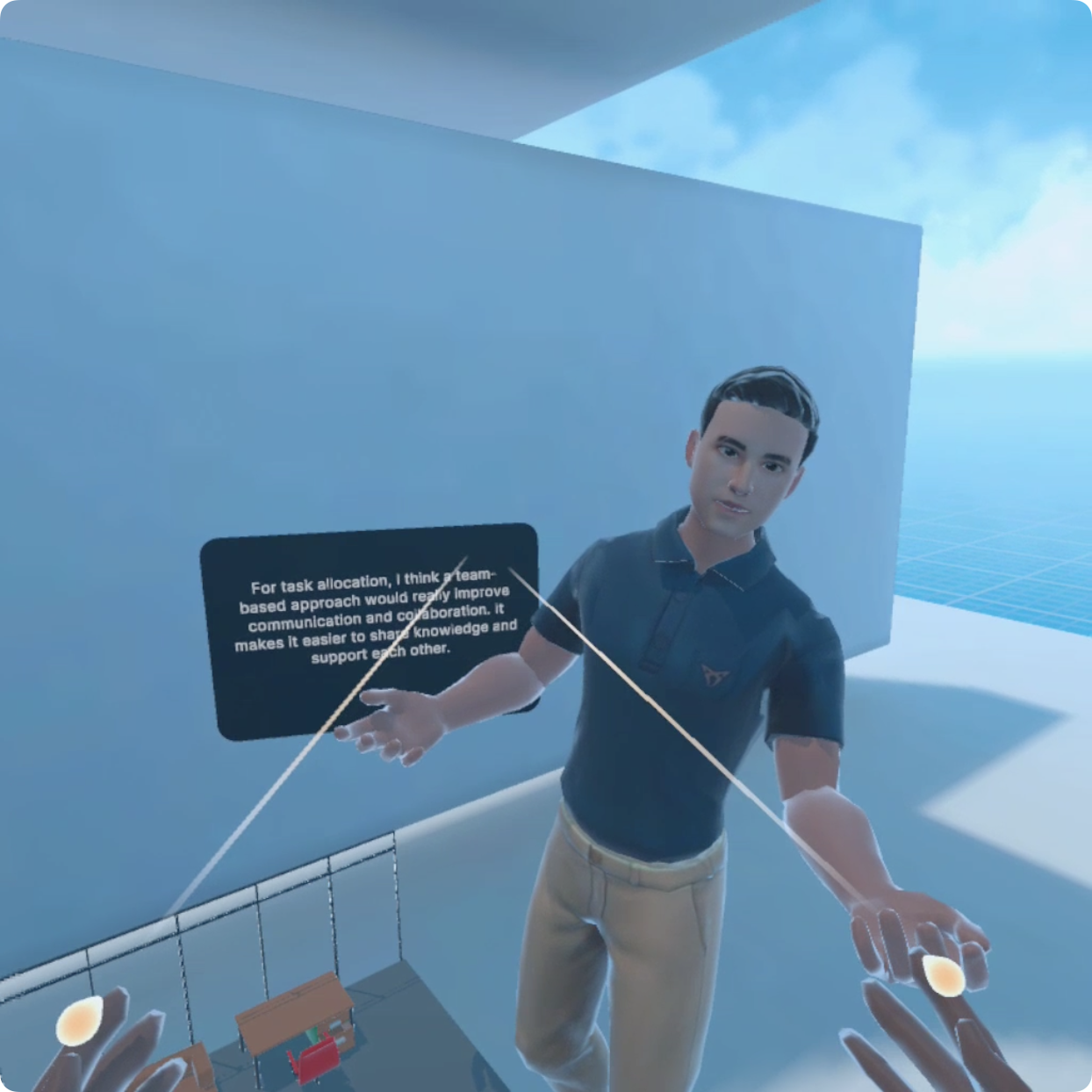}
        \caption{Stand-in responds to attendee's question}
        \label{fig:stand-in-3}
    \end{subfigure}
    \caption{Stand-in behaviours during the meeting. (a) The stand-in begins by listening and observing the ongoing discussion. (b) When a topic aligns with the absentee’s predefined interests or notes, the stand-in actively contributes to the conversation. (c) The stand-in responds to questions posed by other attendees, aligning with the absentee's predefined instructions.}
    \label{fig:stand-in-meeting}
\end{figure*}

In this LLM-powered stand-in system, we utilise the Ready Player Me XR avatar, integrated with RPM SDK v7.3.0, Meta Movement SDK v77, and OpenXR SDK v1.14.1, all synchronised via Fusion v2. The stand-in is programmed to manage the discussion and will contribute input if topics of interest to the absentee are mentioned (see Figure~\ref{fig:stand-in-meeting}). To activate both the meeting and the stand-in, it is necessary to input the meeting ID and the absentee's user ID into the specified editor field. This action ensures the retrieval of the relevant agenda and the absentee's profile details. Assignment of the absentee's RPM avatar prefab in the stand-in field is also required. Once the Unity program is executed within the intended scene, the absentee's stand-in avatar becomes active, and meeting-related information is accessible. Other participants can join using VR headsets once the setup is complete. The stand-in incorporates two animations during the meeting: idle and speaking, sourced from Mixamo. The stand-in will orient towards the current speaker and execute the speaking animation as it interacts.

\begin{figure*}[ht]
    \centering
    \begin{subfigure}[t]{0.45\textwidth}
        \centering
        \includegraphics[width=\textwidth]{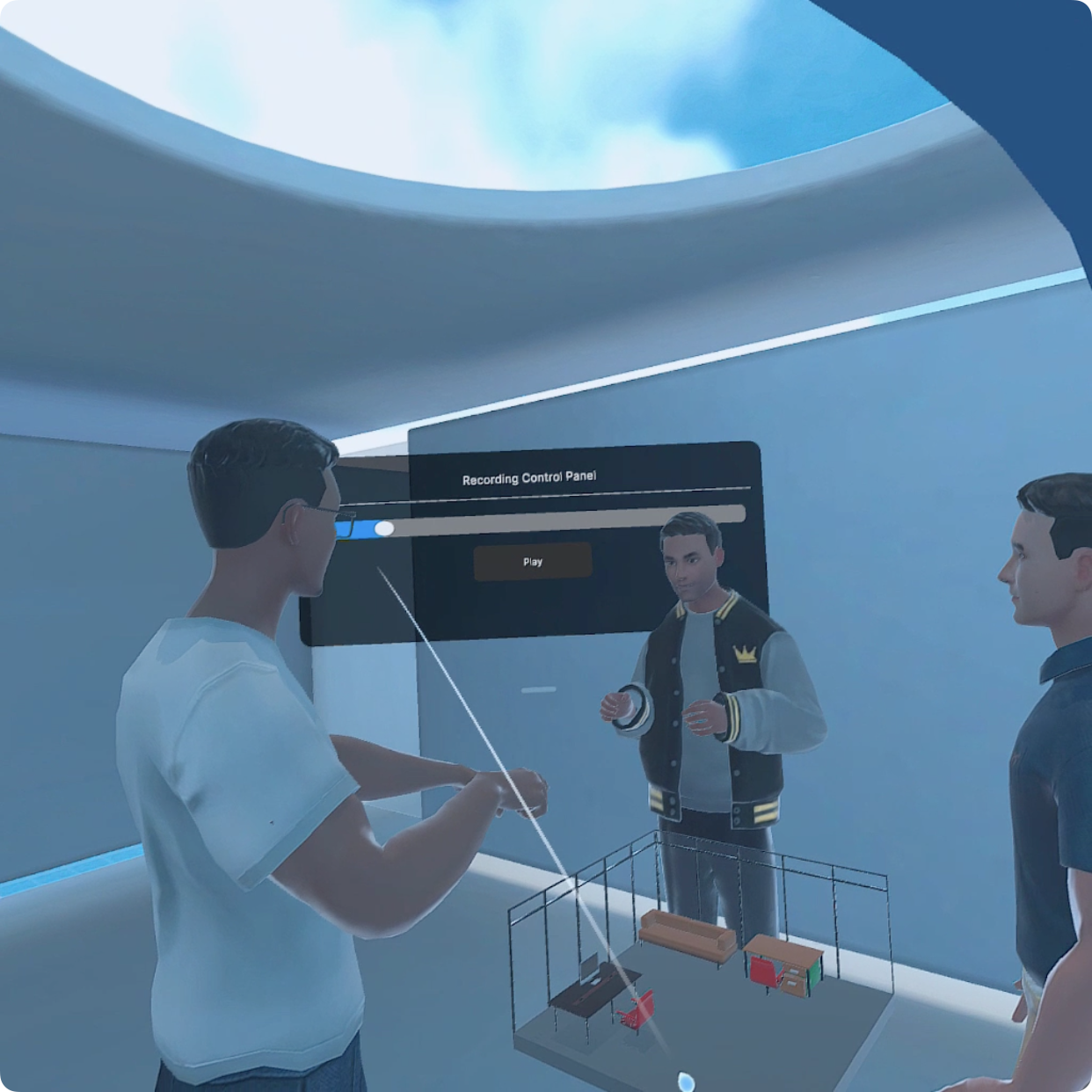}
        \label{fig:stand-in-rec-1}
    \end{subfigure}
    \hfill
    \begin{subfigure}[t]{0.45\textwidth}
        \centering
        \includegraphics[width=\textwidth]{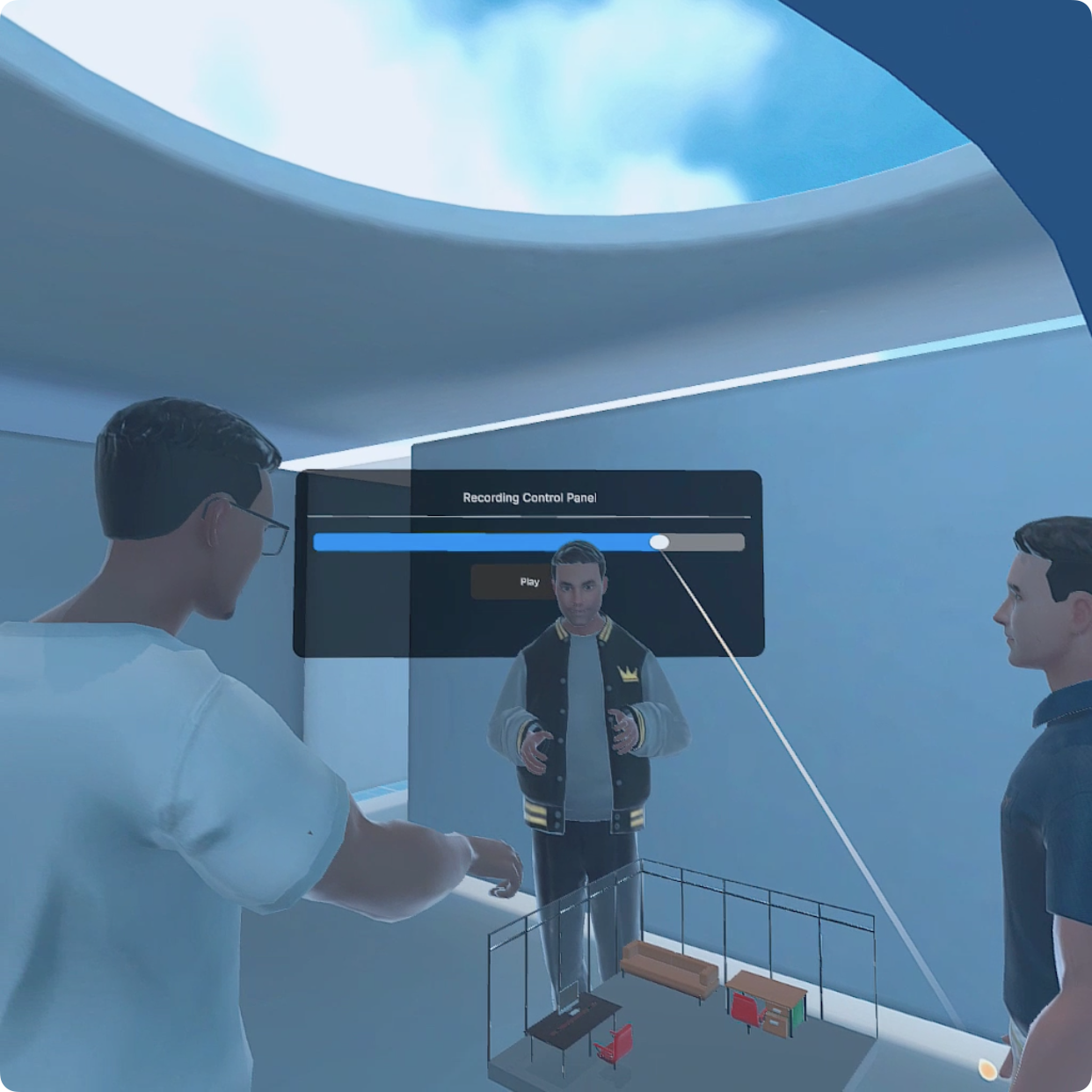}
        \label{fig:stand-in-rec-2}
    \end{subfigure}
    \caption{VR Recording Playback: Users can control the playback timeline using controllers, similar to scrubbing through a YouTube video. All attendee's' movement animations and audio are recorded. Users can freely navigate to any point in time and replay the recorded scene interactively.}
    \label{fig:playback-scene}
\end{figure*}

\subsection{Recording and playback}
We have incorporated the Animation Recorder plugin v1.4.1 and made adjustments to the source code to enable the recording and playback of actions for multiple participants and stand-ins. At the commencement of the meeting, users can initiate the VR meeting recording by manually clicking the "Recording" button in the editor. When recording is halted, all participants can exit the meeting, and the users can terminate the Unity program. The recording files, which contain metadata, store paths for recorded movement data files and audio, including all participants' movement data and the meeting's audio. To replay the meeting, one must switch to the playback scene and enter the metadata file in the specified editor field. The absentee can then wear a VR headset connected to the PC running Unity via Quest Link. Upon accessing Quest Link, the user can start the Playback Scene, and freely navigate using the controllers along the scrubbing timeline (see Figure~\ref{fig:playback-scene}).
 
\section{Limitations and future work}
We identified several challenges in this study. Unlike recording on a 2D screen, VR/MR recordings require the system to record every movement (including gestures, facial expressions, locomotion, etc.), voice and every object's changes in 3D space, which can significantly increase the need for storage for large data and computation power for streaming these changes across users. Additionally, it is also challenging to achieve perfect synchronisation between recorded audio and 3D movements for VR/MR recordings. Advanced algorithms from current 2D video recordings might be helpful in developing the mechanism for synchronising the audio and movement in VR/MR recordings. 

In addition to the technical challenges, privacy concerns beyond the aspect of AI learning from personal information are also a crucial part that needs to be addressed. Given the nature of SEAM, every Iteration is recorded for future audiences to watch. Obtaining the user's consent is essential for a real-world application as the system will record the user's movement and voice. An additional protocol may be provided to the users to clearly understand when the recording starts, where the recording will be stored, and who will have access to this recording. All these processes should be transparent as to how recording is handled in existing videoconferencing platforms to mitigate users' concerns about privacy issues regarding recordings. These challenges suggest that robust recording infrastructure and clear privacy protocol will be necessary before SEAM can be deployed in real-world settings.

We also highlight a few methodological limitations. Although the user studies were conducted with gender balance, Study 1's task only tested with a male stand-in, and it is unclear whether the gender of the stand-in influenced participants' perceptions. In addition, most of our participants were students recruited from our university who lacked experience in workplace meetings. Future work should try SEAM with more professionals who have extensive experience in meetings. Additionally, this research evaluated SEAM through a Wizard-of-Oz (WoZ) methodology in VR, wherein a researcher controlled the behaviour of the stand-in to simulate intelligent interactions. Although this approach enabled an examination of user expectations for optimal engagements, it does not to reflect the constraints, unpredictabilities, and response delays inherent in real-world AI systems. Future work may focus on applying SEAM in other recordable spatial media (e.g. MR) and evaluating AI-powered stand-in in more complicated meeting scenarios.

Further, though we chose a simple meeting topic that any participant could relate to and be able to engage in with minimal training, this might not be representative of the more realistic use cases of a system like this. Thus, it is unclear whether our findings generalise to other meeting scenarios, such as formal meetings with supervisors or colleagues. Future work could focus on different meeting settings to explore how meeting scenarios would affect the interaction with the stand-in and whether participants have different preferences for the stand-in in a dynamic context. Additionally, in Study 1, participants performed the task in the same room together, which may have affected their perception of social presence. It is unclear if a hybrid meeting with more users (e.g. two co-located, two remote) setting would influence the experience of using SEAM. 

Another direction for future research is to explore the granularity of asynchronous participation. In this work, we only tested the scenario where the absentee in Iteration 1 missed the entire meeting. However,  if the absentee can partially participate in the meeting in a different form (e.g., by mobile phone) to provide answers to their stand-in, it is also unclear how this interaction should be established. Thus, future work could focus on exploring the user experience of absentee's partial participation in SEAM. 

\section{Conclusion}
In this paper, we propose a vision for and explore the user experience of Stand-in Enhanced Asynchronous Meetings (SEAM)---meetings that take place across time in which embodied and interactive virtual agents attend the meeting on absentee's behalf, and absentees can contribute to the meeting later by watching the recording from their stand-in's perspective as if they had been there. 

Our results suggest that stand-ins show potential for enabling enhanced meetings that benefit both attendees and absentees. Attendees can involve absentees in the first place by addressing the absentee's stand-in for accessing necessary information that pushes decision-making in the meeting. Absentees later watch the recorded meeting from their stand-in's perspective and feel included as they participated in the original meeting. Our study demonstrates a proof of concept for future asynchronous meetings in which attendees and absentees can interact conversationally more akin to participating in a synchronous meeting. However, we also identified the challenges that lie with SEAMs, such as concerns for trust, privacy, and ethics. We have provided the key design considerations regarding future system improvement and discussed challenges that need to be overcome. In conclusion, this research offers a series of potential opportunities towards meeting systems that support asynchronous collaboration.

\bibliographystyle{ACM-Reference-Format}
\bibliography{references}

\newpage
\appendix
\section{Regression tables}
\begin{table}[h]
\centering
    \caption{Summary of the cumulative probit model for perception of absentee's contribution: Contribution $\sim$ State + (1|PID) + (1|QID). The model includes random intercepts for PID and QID. We provide the posterior means of parameter estimates (Est.), standard errors (Est. Error), and the bounds of the 95\% compatibility interval (CI). All parameter estimates converged with an ESS well above 1000 and an R-hat of 1.00.}
    \Description{This table presents a summary of a cumulative probit model analyzing the perception of an absentee's contribution. It includes ordinal regression cut-points (τ[1] through τ[6]), fixed effects for State (Before), and random effects for Participant (SD). For each parameter, the table provides the estimate with its standard error, and the 95\% confidence interval.}
\begin{tabular}{lll}
    \toprule
    \textbf{Parameter} & \textbf{Estimate (Est. Error)} & \textbf{95\% CI} \\
    \midrule
    \multicolumn{3}{l}{\textbf{Ordinal regression cut-points}} \\
    $\tau$[1] & -3.28 (0.82) & [-5.21, -1.99] \\
    $\tau$[2] & -1.33 (0.39) & [-2.15, -0.61] \\
    $\tau$[3] & -0.35 (0.34) & [-1.01, 0.30] \\
    $\tau$[4] & -0.08 (0.34) & [-0.74, 0.56] \\
    $\tau$[5] & 1.35 (0.39) & [0.66, 2.20] \\
    $\tau$[6] & 3.81 (0.84) & [2.41, 5.64] \\
    \midrule
    \multicolumn{3}{l}{\textbf{Fixed Effects}} \\
    State (Before) & 0.64 (0.30) & [0.04, 1.24] \\
    \midrule
    \multicolumn{3}{l}{\textbf{Random Effects}} \\
    Participant (SD) & 1.28 (0.35) & [0.67, 2.06] \\
    \bottomrule
\end{tabular}
\label{tab:perception-contribution}
\end{table}

\begin{table}[h]
\centering
    \caption{Summary of the cumulative probit model for perception of absentee's inclusion: Inclusion $\sim$ State + (1|PID) + (1|QID). The model includes random intercepts for PID and QID. We provide the posterior means of parameter estimates (Est.), standard errors (Est. Error), and the bounds of the 95\% compatibility interval (CI). All parameter estimates converged with an ESS well above 1000 and an R-hat of 1.00.}
    \Description{Table 3 presents a summary of a cumulative probit model that analyzes the perception of an absentee’s inclusion, with the model formula being Inclusion ~ State + (1|PID) + (1|QID). The table lists the parameter estimates, their standard errors, and the 95\% compatibility intervals for both fixed and random effects. The table emphasizes that all parameter estimates converged with an effective sample size (ESS) well above 1000 and an R-hat of 1.00.}
\begin{tabular}{lll}
    \toprule
    \textbf{Parameter} & \textbf{Estimate (Est. Error)} & \textbf{95\% CI} \\
    \midrule
    \multicolumn{3}{l}{\textbf{Regression Coefficients:}} \\
    $\tau$[1] & -3.62 (0.72) & [-5.13, -2.37] \\
    $\tau$[2] & -2.60 (0.57) & [-3.78, -1.59] \\
    $\tau$[3] & -1.65 (0.46) & [-2.58, -0.82] \\
    $\tau$[4] & -1.13 (0.42) & [-1.99, -0.34] \\
    $\tau$[5] & 0.55 (0.40) & [-0.20, 1.38] \\
    $\tau$[6] & 2.56 (0.57) & [1.56, 3.80] \\
    \midrule
    \multicolumn{3}{l}{\textbf{Fixed Effects}} \\
    State (Before)& -0.33 (0.28) & [-0.89, 0.22]  \\
   
    \midrule
    \multicolumn{3}{l}{\textbf{Random Effects}} \\
    PID (SD) & 1.58 (0.41) & [0.90, 2.48] \\
    \bottomrule
\end{tabular}
\label{tab:perception-inclusion}
\end{table}

\begin{table}[h]
\centering
    \caption{Summary of the cumulative probit model for perception of respectful behaviours: Behaviours $\sim$ State + (1|PID) + (1|QID). The model includes random intercepts for PID and QID. We provide the posterior means of parameter estimates (Est.), standard errors (Est. Error), and the bounds of the 95\% compatibility interval (CI). All parameter estimates converged with an ESS well above 1000 and an R-hat of 1.00.}
    \Description{This table, labeled as "Table 4," summarizes the results of a cumulative probit model analyzing the perception of respectful behaviors. The model includes random intercepts for PID and QID. It reports on the posterior means of parameter estimates (with their estimated errors), and the bounds of the 95\% compatibility interval (CI) for each parameter. The regression coefficients (τ[1] to τ[5]) show varying estimates along with their standard errors and 95\% CIs, indicating the influence of different parameters on the model, some of which include negative and positive values. Additionally, the table displays fixed effects related to the state (before) and random effects, specifically the standard deviation (SD) for PID. The relevant statistics indicate that all parameter estimates converged effectively with a high effective sample size (ESS), which is well above 1000 and an R-hat of 1.00.}
\begin{tabular}{lll}
    \toprule
    \textbf{Parameter} & \textbf{Estimate (Est. Error)} & \textbf{95\% CI} \\
    \midrule
    \multicolumn{3}{l}{\textbf{Regression Coefficients:}} \\
    $\tau$[1] & -2.90 (0.61) & [-4.19, -1.81] \\
    $\tau$[2] & -1.77 (0.45) & [-2.71, -0.94] \\
    $\tau$[3] & -1.21 (0.40) & [-2.04, -0.49] \\
    $\tau$[4] & -0.17 (0.36) & [-0.89, 0.55] \\
    $\tau$[5] & 2.21 (0.53) & [1.31, 3.39] \\
    \midrule
    \multicolumn{3}{l}{\textbf{Fixed Effects}} \\
    State (Before)  & 0.49 (0.32) & [-0.11, 1.13] \\
   
    \midrule
    \multicolumn{3}{l}{\textbf{Random Effects}} \\
    PID (SD) & 1.44 (0.38) & [0.78, 2.27] \\
    \bottomrule
\end{tabular}
\label{tab:perception-behaviours}
\end{table} 
\break

\begin{table*}[!h]
    \centering
    \caption{Summary of the cumulative probit model for perception of the attendee and absentee's social presence in Iteration 1 and Iteration 2: Social presence $\sim$ Attendance + (1|Participant) + (1|Questions). We provide the posterior means of parameter estimates (Estimate), standard errors (Error), and the bounds of the 95\% compatibility interval (CI). All parameter estimates converged with an ESS well above 1000 and an R-hat of 1.00. Fixed effects are relative to the perceptions of the attendee in iteration 1.}
    \Description{This table presents a summary of a cumulative probit model analyzing the perception of social presence for attendees and absentees in a study. It displays parameter estimates, standard errors, and 95\% confidence intervals for two categories: "Perception of self" and "Perception of the other." The table is divided into three sections: Ordinal regression cut-points (τ[1] to τ[6]), Fixed Effects (Attendance), and Random Effects (Participant and Question). Each row shows the estimate, standard error, and confidence interval for the respective parameter.}
    \label{tab:perception-social-presence_full}
    \begin{tabularx}{\textwidth}{>{\raggedright\arraybackslash}p{3cm}XXXX}
    \toprule
    \multirow{2}{*}{\textbf{Parameter}} & \multicolumn{2}{c}{\textbf{Perception of self}} & \multicolumn{2}{c}{\textbf{Perception of the other}} \\
    \cmidrule(lr){2-3} \cmidrule(lr){4-5}
    & \textbf{Estimate (Error)} & \textbf{95\% CI} & \textbf{Estimate (Error)} & \textbf{95\% CI} \\ \midrule
    \multicolumn{5}{l}{\textbf{Ordinal regression cut-points}} \\
    $\tau$[1]  &   -2.67 (0.22) & [-3.09, -2.24]   &  -3.16 (0.25) & [-3.65, -2.68] \\
    $\tau$[2]  &   -1.78 (0.21) & [-2.18, -1.36]   &  -2.15 (0.23) & [-2.62, -1.69] \\
    $\tau$[3]  &   -1.21 (0.21) & [-1.60, -0.80]   &  -1.48 (0.23) & [-1.94, -1.03] \\
    $\tau$[4]  &   -0.79 (0.21) & [-1.19, -0.38]    &  -0.96 (0.23) & [-1.41, -0.52]  \\
    $\tau$[5]  &   -0.07 (0.21) & [-0.47, 0.33]     &  -0.14 (0.23)  & [-0.61, 0.29]    \\
    $\tau$[6]  &    1.01 (0.21) & [0.61, 1.42]     &  1.06 (0.23)  & [0.59, 1.51]    \\ \midrule 
    \multicolumn{5}{l}{\textbf{Fixed Effects}} \\
    Absentee in Iteration 1  & -0.75 (0.07) & [-0.97, -0.61] & -0.73 (0.07) & [-0.86, -0.60] \\ 
    Absentees in Iteration 2 & -0.16 (0.17) & [-0.48, 0.18] & -0.54 (0.25) & [-1.03, -0.04] \\  \midrule
    \multicolumn{5}{l}{\textbf{Random Effects}} \\
    Participant (SD) & 0.47 (0.07) & [0.35, 0.62] & 0.73 (0.09) & [0.57, 0.94]  \\
    Question (SD) & 0.61 (0.12) & [0.42, 0.90]& 0.53 (0.11) & [0.36, 0.78] \\
    Meeting (SD) & 0.34 (0.10) & [0.19, 0.58] & 0.46 (0.13) & [0.27, 0.77] \\
    \bottomrule
    \end{tabularx}
\end{table*}
\break

\section{Questionnaire Items}
\begin{table*}[h]
\centering
\caption{Self-designed questions}
\Description{From Table 5, the self-designed questions are:
1. I felt Lee made a meaningful contribution to the decision-making process
2. I felt Lee was included in this meeting
3. I felt my behaviours towards Lee were respectful.}
\begin{tabular}{@{}lll@{}}
\toprule
Self-designed questions\\
\midrule
I felt Lee made a meaningful contribution to the decision-making \\
process. \\
I felt Lee was included in this meeting. \\
I felt my behaviours towards Lee were respectful.\\
\bottomrule
\end{tabular}
\label{tab:self-designed-questions}
\end{table*}


\section{Interview questions}

\begin{table*}[h]
\centering
\caption{Study 1 \& 2 first interview questions}
\Description{From Table 6, the first interview questions are:
1. How many people were in the meeting?
2. Why did they miss the meeting?
3. When did they inform other members that they couldn’t come?
4. What was the experience of that meeting?
5. How did this missing member catch up?
6. How did the absent member contribute to the meeting?}
\begin{tabular*}{\textwidth}{@{\extracolsep{\fill}}lll@{}}
\toprule
Structured questions \\
\midrule
How many people were in the meeting? \\
Why did they miss the meeting? \\
When did they inform other members that they couldn’t come? \\
What was the experience of that meeting? \\
How did this missing member catch up? \\
How did the absent member contribute to the meeting?\\
\bottomrule
\end{tabular*}
\label{tab:first-interview}
\end{table*} 

\begin{table*}[h]
\centering
\caption{Study 1 second interview questions}
\Description{From Table 7, the second interview questions are:
How did you feel about having Lee’s stand-in in the meeting?
How did you feel when addressing Lee (stand-in) during the meeting?
Compared to typical meetings (Zoom or face-to-face) where someone cannot attend, how did the presence of the stand-in influence the way you behaved during this VR meeting?
Did you notice any differences between interacting with your partner and the stand-in during the meeting?
Can you explain these differences in more detail?
How many people did you perceive, 3 or 2? Why?}
\begin{tabular}{@{}lll@{}}
\toprule
Structured questions\\
\midrule
How did you feel about having Lee's stand-in in the meeting? \\
How did you feel when addressing Lee (stand-in) during the meeting? \\
Compared to typical meetings (Zoom or face-to-face) where someone cannot attend, how did the presence of the stand-in influence \\ the way you behaved during this VR meeting? \\
Did you notice any differences between interacting with your partner and the stand-in during the meeting?\\
Can you explain these differences in more detail? \\
How many people did you perceive, 3 or 2? Why? \\
\bottomrule
\end{tabular}
\label{tab:second-interview}
\end{table*} 

\begin{table*}[h]
\centering
\caption{Study 1 third interview questions}
\Description{Here are the third interview questions listed in Table 8:
1. What was the experience of watching the recording from an absent attendee's perspective?
2. How did you feel when being addressed as an absent attendee?
3. After reviewing your behaviours in the recording, would you change anything about your behaviors next time with a stand-in? Why?
4. How does the experience of reviewing this VR recording differ from catching up on a missed meeting via traditional methods?
5. Will you watch the recording like this? Why？
6. After trying the system from both present and absent attendee's perspective, can you describe what was the overall experience?
7. Which parts of those two VR activities were impressive, either positive or negative?
8. How would you expect your stand-in to represent you?
9. What factors will you consider?
10. How would you want this stand-in to behave and communicate on your behalf?
11. How will you design your stand-in's behaviors?
12. Which avatar would you choose to be your stand-in? (Avaturn, ReadyPlayerMe, Meta Avatar)
13. In what other situations or scenarios, would you use such stand-in?
14. Do you have any other questions or comments?}
\begin{tabular}{@{}lll@{}}
\toprule
Structured questions\\
\midrule
What was the experience of watching the recording from an absent attendee’s perspective? \\
How did you feel when being addressed as an absent attendee? \\
After reviewing your behaviours in the recording, would you change anything about your behaviors next time with a stand-in? Why? \\
How does the experience of reviewing this VR recording differ from catching up on a missed meeting via traditional methods? \\
Will you watch the recording like this? Why? \\
After trying the system from both present and absent attendee’s perspective, can you describe what was the overall experience? \\
Which parts of those two VR activities were impressive, either positive or negative? \\
How would you expect your stand-in to represent you?\\ 
What factors will you consider? \\
How would you want this stand-in to behave and communicate on your behalf? \\
How will you design your stand-in's behaviors? \\
Which avatar would you choose to be your stand-in? (Avaturn, ReadyPlayerMe, Meta Avatar) \\
In what other situations or scenarios, would you use such stand-in? \\
Do you have any other questions or comments? \\
\bottomrule
\end{tabular}
\label{tab:third-interview}
\end{table*} 

\begin{table*}[h]
\centering
\caption{Study 2 second interview questions}
\Description{Here are the third interview questions listed in Table 8:
1. What was the experience of watching the recording from an absent attendee's perspective?
2. What factors or things made you feel included?
3. What factors or things made you feel not included?
4. How did you feel about seeing your stand-in’s response?
5. What made you decide to adjust response or add comments (if any)?
6. How did you feel when being addressed as an absent attendee?
7. How does the experience of reviewing this VR recording differ from catching up on a missed meeting via traditional methods?
8. In which situations you would watch the recording like this (if any)? Why?
9. How would you expect your stand-in to represent you?
10. What factors will you consider?
11. How would you want this stand-in to behave and communicate on your behalf?
12. How will you design your stand-in's behaviors?
13. Which avatar would you choose to be your stand-in? (Avaturn, ReadyPlayerMe, Meta Avatar)
14. Do you have any other questions or comments?}
\begin{tabular}{@{}lll@{}}
\toprule
Structured questions\\
\midrule
What was the experience of watching the recording from an absent attendee's perspective? \\
What factors or things made you feel included? \\
What factors or things made you feel not included? \\
How did you feel about seeing your stand-in’s response? \\
What made you decide to adjust response or add comments (if any)? \\
How did you feel when being addressed as an absent attendee? \\
How does the experience of reviewing this VR recording differ from catching up on a missed meeting via traditional methods? \\
In which situations you would watch the recording like this (if any)? Why? \\
How would you expect your stand-in to represent you? \\
What factors will you consider? \\
How would you want this stand-in to behave and communicate on your behalf? \\
How will you design your stand-in's behaviors? \\
Which avatar would you choose to be your stand-in? (Avaturn, ReadyPlayerMe, Meta Avatar) \\
Do you have any other questions or comments? \\
\bottomrule
\end{tabular}
\label{tab:study-2-interview-2}
\end{table*}


\end{document}